\documentclass[epsfig,useAMS,useAMSmath,usenatbib]{mn2e}
\usepackage{enumerate,color,ulem}

\usepackage{url}
\usepackage{epsfig,epstopdf}

\newcommand{\HI}{H{\sc i}}
\newcommand{\kms}{\ensuremath{\mathrm{km\,s^{-1}}}}
\newcommand{\df}{{\sc DiskFit}}
\newcommand{\vf}{{\sc velfit}}
\newcommand{\FITS}{{\footnotesize FITS}}

\title[Non-Axisymmetries in NGC~6503]{Searching for Non-axisymmetries in NGC~6503: A Weak End-on Bar}

\author[Kuzio de Naray et al.]
{Rachel Kuzio de Naray$^{1}$\thanks{Email: kuzio@rmc.ca (RKD); Cameron.Arsenault@forces.gc.ca (CAA); Kristine.Spekkens@rmc.ca (KS); sellwood@physics.rutgers.edu (JAS); mcdonald@space.mit.edu (MM); jsimon@obs.carnegiescience.edu (JDS); teuben@astro.umd.edu (PT)}\thanks{Visiting Astronomer, Kitt Peak National Observatory, National Optical Astronomy Observatory, which is operated by the Association of Universities for Research in Astronomy, Inc. (AURA) under cooperative agreement with the National Science Foundation.}, Cameron A.~Arsenault$^{1}$\footnotemark[1], Kristine Spekkens$^{1}$\footnotemark[1],\newauthor J.A.~Sellwood$^{2}$\footnotemark[1], Michael McDonald$^{3}$\footnotemark[1], Joshua D.~Simon$^{4}$\footnotemark[1]\footnotemark[2] and Peter Teuben$^{5}$\footnotemark[1]\\
$^{1}$Department of Physics, Royal Military College of Canada, P.O. Box 17000, Station Forces, Kingston, ON, K7K 7B4, Canada\\
$^{2}$Department of Physics \& Astronomy, Rutgers, The State University of New Jersey, 136 Frelinghuysen Road, Piscataway, NJ 08854-8019, USA\\
$^{3}$Kavli Institute for Astrophysics and Space Research, MIT, Cambridge, MA 02139, USA\\
$^{4}$Observatories of the Carnegie Institution of Washington, 813 Santa Barbara Street, Pasadena, CA 91101, USA\\
$^{5}$Department of Astronomy, University of Maryland, College Park, MD 20742, USA}

\begin{document}

\pagerange{\pageref{firstpage}--\pageref{lastpage}} \pubyear{Accepted to MNRAS 2012 September 13}

\maketitle

%%%%%%%%%%%%%%%%%%%%%%%%%%%%
\begin{abstract}
%%%%%%%%%%%%%%%%%%%%%%%%%%%%
The isolation, simple apparent structure, and low luminosity of the nearby spiral galaxy NGC~6503 make it an ideal candidate for an in-depth kinematic and photometric study.  We introduce a new publicly available code, \df, that implements procedures for fitting non-axisymmetries in either kinematic or photometric data.  We use \df\ to analyze new H$\alpha$ and CO velocity field data as well as \HI\ kinematics from Greisen et al.\ to search for non-circular motions in the disc of NGC~6503.  We find NGC~6503 to have remarkably regular gas kinematics that are well-described by rotation only.  We also use \df\ and a new $K$s-band image of NGC~6503 to constrain photometric models of the disc, bar and bulge.  We find the galaxy to be photometrically dominated by the disc.  We find NGC~6503 to contain a faint bar and an exponential bulge which together contribute only $\sim$~5\% of the total galaxy light.  The combination of our kinematic and photometric \df\ models suggest that NGC~6503 contains a weak, end-on bar that may have produced its Type II surface brightness profile but is unlikely to be responsible for its strong $\sigma$-drop.
\end{abstract}

%%%%%%%%%%%%%%%%%%%%%%%%%%%%
%%%%%%%%%%%%%%%%%%%%%%%%%%%%

\begin{figure*}
\includegraphics[scale=0.30]{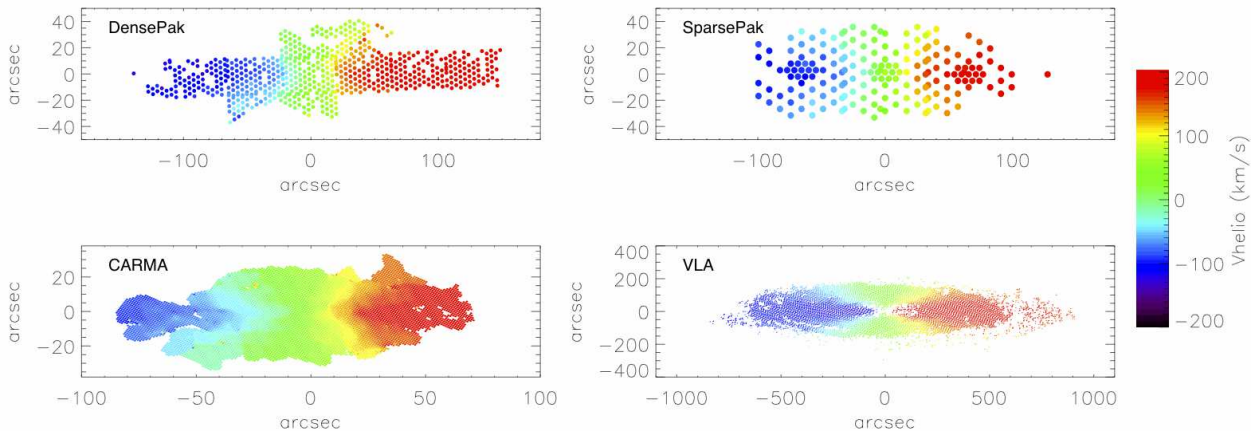}
\caption{Observed velocity fields of NGC~6503.  All colourscales are identical, and all velocity fields have been rotated 30\degr\ to the west.  Note the spatial extent is different in each panel.  1$\arcsec$~$\approx$~0.023~kpc \citep{Mould00}.  Top left: DensePak H$\alpha$ data.  The diameter of the DensePak fibres is 3$\arcsec$.  Top right: SparsePak H$\alpha$ data.  The SparsePak fibres are 5$\arcsec$ in diameter.  Bottom left: CO data.  The spatial resolution is 4.2$\arcsec$.  Note the smaller spatial extent of the CO data compared to the H$\alpha$ data.  Bottom right: \HI\ velocity field derived using the Modified Envelope Tracing technique of \citet{Gentile04} on the \HI\ data cube from \citet{Greisen09}.  The spatial resolution is 14$\arcsec$.  Note that the spatial extent of the \HI\ velocity field is about 7 times larger than the optical H$\alpha$ data. (A colour version of this figure is available in the online journal.) \label{allvfs}}
\end{figure*}

%%%%%%%%%%%%%%%%%%%%%%%%%%%%
\section{Introduction}
%%%%%%%%%%%%%%%%%%%%%%%%%%%%
\label{intro}
NGC~6503 is a low-luminosity, late-type spiral galaxy that is relatively isolated from systems of similar mass.  It has been the focus of several recent studies to characterize its underlying disc structure.   Using Very Large Array (VLA\footnote{The VLA is operated by the National Radio Astronomy Observatory, which is a facility of the National Science Foundation, operated under cooperative agreement by Associated Universities, IIlC.}) observations, \citet*{Greisen09} find the kinematics of the \HI\ disc to be remarkably regular.  They also find that the data require a disc with both thin and thick components.  \citet*{Puglielli10} construct dynamical models of NGC~6503 using Bayesian statistics and Markov chain Monte Carlo (MCMC) techniques.  Their extensive modeling places interesting constraints on the structure of the galaxy disc and bulge, mass-to-light ratios, and the dark matter halo.  They find the bulge of the galaxy to have a low S\'ersic index and a mass-to-light ratio lower than the disc, and that the dark matter halo is cuspy.

\citet*[][hereafter F10]{Freeland10} present WIYN\footnote{Based on observations obtained at the WIYN Observatory.  The WIYN Observatory is a joint facility of the University of Wisconsin-Madison, Indiana University, Yale University, and the National Optical Astronomy Observatory.} High-resolution Infrared Camera (WHIRC) $H$-band (1.6$\mu$m) imaging, in combination with multi-wavelength archival imaging and data from the literature, to investigate the origin(s) of the observed star-forming ring and central $\sigma$-drop.  They find a plateau in the surface brightness profile (Freeman Type II behaviour, \citealt{Freeman70}) and several structures in the disc in addition to the star-forming ring: an end-on bar, a circumnuclear disc, and a nuclear spiral.   Comparing to the simulations of \citet[][hereafter B05]{Bureau05}, they argue that the bar in NGC~6503 is strong and that it is this structure which can explain the photometric and kinematic features.  In this model, the remarkable regularity of the kinematics in NGC~6503 therefore belies a strong non-axisymmetry near the centre.

\defcitealias{Bureau05}{B05}

Because of its suitability for detailed structural studies, we have acquired new, high-quality, multi-wavelength kinematics and photometry of NGC~6503.  In this paper, we use a newly released code, \df, to perform an in-depth study of the underlying structure of the galaxy.  We investigate the gas kinematics and search for non-circular motions in H$\alpha$, CO, and \HI\ velocity field data.  We also use $K$s-band imaging to constrain photometric models of the disc, bar, and bulge.

In Section~\ref{data}, we present the new observational data for NGC~6503.  In Section~\ref{discfit}, we introduce and describe the first \df\ release, that we use to fit non-parametric models to the galaxy data.  The results of the kinematic modeling are presented in Section~\ref{Kresults}; photometric modeling results are in Section~\ref{Presults}.  In Section~\ref{Bar}, we discuss the kinematic and photometric constraints on the structure of NGC~6503.  We summarize our results in Section~\ref{summary}.

\defcitealias{Freeland10}{F10}

%%%%%%%%%%%%%%%%%%%%%%%%%%%%
\section{Data} 
%%%%%%%%%%%%%%%%%%%%%%%%%%%%
\label{data}
The isolation, simple apparent structure, and low-luminosity of NGC~6503 make it an ideal candidate for  kinematic and photometric study. We therefore compile a multi-wavelength dataset of high-resolution spectroscopy and photometry for the galaxy.  

We obtain new integral field unit (IFU) H$\alpha$ velocity fields using both the DensePak and SparsePak IFUs on the 3.5-m WIYN telescope at Kitt Peak National Observatory (KPNO).  We also obtain CO(J=1-0) observations using the Combined Array for Research in Millimeter-wave Astronomy (CARMA\footnote{Support for CARMA construction was derived from the Gordon and Betty Moore Foundation, the Kenneth T. and Eileen L. Norris Foundation, the James S. McDonnell Foundation, the Associates of the California Institute of Technology, the University of Chicago, the states of California, Illinois, and Maryland, and the National Science Foundation. Ongoing CARMA development and operations are supported by the National Science Foundation under a cooperative agreement, and by the CARMA partner universities. }).  To minimize the effects of beam smearing, we derive a new \HI\ velocity field from the \citet{Greisen09} data cube.  We also obtain new Wide-field Infrared Camera (WIRCAM) $K$s-band imaging using the  3.6-m Canada-France-Hawaii Telescope (CFHT\footnote{Based on observations obtained with WIRCam, a joint project of CFHT, Taiwan, Korea, Canada, France, and the Canada-France-Hawaii Telescope (CFHT) which is operated by the National Research Council (NRC) of Canada, the Institute National des Sciences de l'Univers of the Centre National de la Recherche Scientifique of France, and the University of Hawaii.}).  In the following sections, we present each of the datasets and describe the observations in detail.

%%%%%%%%%%%%%%%%%%%%%%%%%%%%
\subsection{H$\alpha$ velocity fields} 
%%%%%%%%%%%%%%%%%%%%%%%%%%%%
\label{HAdata}
We have obtained H$\alpha$ velocity fields of NGC~6503 using the DensePak \citep{Barden98} and SparsePak \citep{Bershady04} IFUs on the WIYN telescope at KPNO.  The DensePak IFU is comprised of 85 working 3\arcsec\ diameter fibres arranged in a fixed 43\arcsec~$\times$~28\arcsec\ rectangle. Similarly, the SparsePak IFU is comprised of 82 5\arcsec\ diameter fibres arranged in a fixed 70\arcsec~$\times$~70\arcsec\ rectangle.  

The DensePak observations were obtained on 2004 July 30.  The instrument setup matched that of \citet{Simon03} and \citet{Simon05}, with the Bench Spectrograph in echelle mode with 13~km~s$^{-1}$ velocity resolution over a narrow spectral range centred on H$\alpha$.  The galaxy was observed with 11 DensePak pointings, seven along the major axis, two on the minor axis, and one each to the northwest and southeast, providing 654 independent velocity measurements across the disc.  Because the H$\alpha$ emission was quite bright, most of the pointings received a single 1200~s exposure, with a second matching exposure at a few positions where the emission was fainter.

\begin{figure}
\center
\includegraphics[scale=0.43]{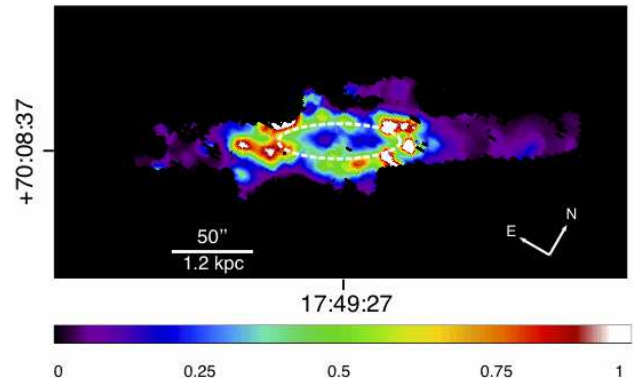}
\caption{H$\alpha$ intensity map from the DensePak observations.  The dashed white ellipse indicates the position of the star-forming ring discussed by \citetalias{Freeland10} and in Section~\ref{Features}.  The linear flux scaling is arbitrary. (A colour version of this figure is available in the online journal.) \label{dpharing}}
\end{figure}

Data reduction followed the outline of \citet{Simon03} and \citet{Simon05}.  The data were reduced in IRAF\footnote{IRAF is distributed by the National Optical Astronomy Observatory, which is operated by the Association of Universities for Research in Astronomy (AURA), Inc., under agreement with the National Science Foundation.} using the \textsc{hydra} package.  The individual frames were bias-subtracted and flat fielded, and then cosmic rays were removed.  The spectrum of each fibre was extracted and wavelength calibrated by comparison with a ThAr lamp frame.  Next, we averaged together the four sky fibres, leaving out any sky spectra that were contaminated by emission lines from NGC~6503.  We then removed a linear baseline, performed a Gaussian fit to the averaged sky emission near H$\alpha$, and subtracted the fit from all of the data fibres.  Finally, individual exposures of the same field were averaged together when present, and velocities were measured for each fibre by fitting a Gaussian to the observed H$\alpha$ emission line.  The median velocity uncertainty from the Gaussian fits was 1.2~km~s$^{-1}$.  The observed H$\alpha$ intensities in each pointing were cross-correlated against a narrow-band H$\alpha$ image of the galaxy from \citet{Kennicutt08} to determine the absolute position of each observation (see \citealt{Simon03}).  The observed DensePak H$\alpha$ velocity field is shown in Fig.~\ref{allvfs}.  A map of the H$\alpha$ intensity in the DensePak fibres is shown in Fig.~\ref{dpharing}.  The star-forming ring discussed by \citetalias{Freeland10} is visible in the H$\alpha$ intensity map.

NGC~6503 was observed with the SparsePak IFU on 2009 May 16.  The STA1 CCD was used with the 316@63.4 grating in eighth order, centred near H$\alpha$, giving a 40~\kms\ velocity resolution.  The SparsePak array was aligned with the major axis of the galaxy and three pointings were used to cover the length of the galaxy.  Individual exposures were 1200~s, and two exposures were taken at each pointing.  A ThAr lamp was observed to provide wavelength calibration.

The SparsePak data were reduced in IRAF using the \textsc{hydra} package following the procedure outlined in \citet{Kuzio06}.  The data were bias-subtracted and flattened, and the IRAF task \texttt{dohydra} was used to extract the spectra.  The spectra were wavelength-calibrated using a wavelength solution created from the observations of the ThAr lamp.  The two exposures per pointing were combined to remove cosmic rays and to increase the signal-to-noise.  Sky subtraction was not performed, as the night-sky emission lines were used as the reference wavelengths \citep{Osterbrock96} by which the velocities of the galactic emission lines were measured.   

The observed fibre velocities were measured by fitting Gaussians to both the sky lines and the five galactic emission lines of interest: H$\alpha$, [\mbox{N\,{\sc ii}}]$\lambda$6548, [\mbox{N\,{\sc ii}}]$\lambda$6584, [\mbox{S\,{\sc ii}}]$\lambda$6717, and [\mbox{S\,{\sc ii}}]$\lambda$6731.  There was less scatter between the measured galactic emission line velocities when using the night-sky calibration than when using the ThAr calibration.  The arithmetic mean of the measured emission-line velocities in each fibre was used as the fibre velocity and the error on the fibre velocity was set to the maximum difference between the measured velocities and the mean.  Most of the errors are less than 5~\kms, though a few were greater than $\sim$10~\kms.  The observed SparsePak H$\alpha$ velocity field is shown in Fig.~\ref{allvfs}.

\begin{figure}
\center
\includegraphics[scale=0.43]{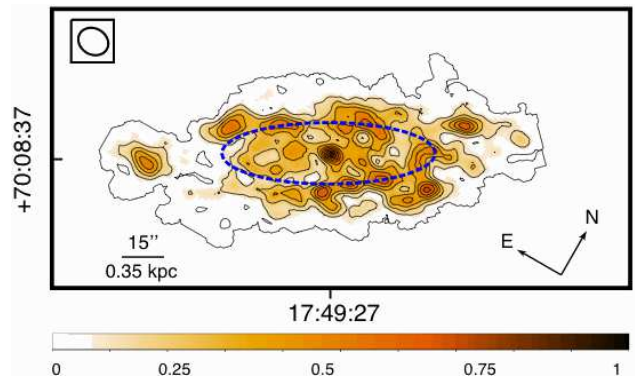}
\caption{CARMA CO (J=1-0) intensity map. Contour levels are at evenly spaced intervals from 10\% to 90\% of the peak intensity.  The blue dashed ellipse indicates the position of the star-forming ring discussed by \citetalias{Freeland10} and in Section~\ref{Features}.  The beam size is plotted in the top left corner. (A colour version of this figure is available in the online journal.) \label{COmom0}}
\end{figure}

%%%%%%%%%%%%%%%%%%%%%%%%%%%%
\subsection{CO velocity field} 
%%%%%%%%%%%%%%%%%%%%%%%%%%%%
\label{COdata}
NGC~6503 was observed with the 15-antenna (10-m and 6-m dishes) CARMA array in March/April 2008 and October/November 2009.  The instrument setup, data reduction and calibration are the same as those described in \citet{Koda11}.  We used 1642+689 or 1849+670 as phase calibrator, interrupting source integrations every 20 minutes.  Absolute flux calibration was done using MARS, NEPTUNE or MWC349, whichever was available, and 3C273/3C345 as bandpass calibrator.  We mosaicked the galaxy with 9 pointings: 5 beams along the major axis, and 2 beams on either side of the minor axis.  Standard routines from the \textsc{miriad} package \citep{Sault95} were used for calibration and mapping.  A final CO datacube was produced with a spatial resolution of 4.2\arcsec~$\times$~3.4\arcsec, and in velocity 5~\kms.  

The CO velocity field shown in Fig.~\ref{allvfs} is derived from the first moment of the datacube, where only emission exceeding 2.5 mJy/beam after the data were spatially smoothed by 3$\arcsec$ and spectrally smoothed by 15~\kms\ were retained.  Note the smaller spatial extent of the CO data compared to the H$\alpha$ data.  In Fig.~\ref{COmom0} we show the intensity map of the CO data and indicate the position of the star-forming ring discussed by \citetalias{Freeland10}.

%%%%%%%%%%%%%%%%%%%%%%%%%%%%
\subsection{\HI\ velocity field} 
%%%%%%%%%%%%%%%%%%%%%%%%%%%%
\label{HIdata}
The reduced \HI\ datacube for NGC~6503 is from \citet{Greisen09}.  Briefly, these data were obtained using the VLA in C configuration. There were 127 spectral channels, each being separated by 24.4~kHz (5.15~\kms), and centred on 26~\kms\ heliocentric radial velocity.   The total on-source integration time was just over 500~minutes, and the data were imaged to produce an angular resolution of 14$\arcsec$.  For complete details on the observing setup and data reduction, the reader is referred to \citet{Greisen09}.

In order to minimize the effects of beam smearing in this high inclination system (see Table~\ref{datatable}), we use the Modified Envelope Tracing technique of \citet{Gentile04} to derive a velocity field from the \citet{Greisen09} data cube. The resulting velocity field is shown in Fig.~\ref{allvfs}.  Note that the radial extent of the \HI\ data is about 7 times larger than that of the H$\alpha$ data. Beyond $r=300$\arcsec\ where beam smearing is negligible, the velocity field is almost identical to that obtained from the first moment of the \HI\ distribution in figure 8 of \citet{Greisen09}. At $r<300$\arcsec\ where beam smearing produces skewed line-of-sight \HI\ profiles, the projected velocities in Fig.~\ref{allvfs} have larger amplitudes than in the first moment map, by an average of 10~\kms\ along the major axis.

%%%%%%%%%%%%%%%%%%%%%%%%%%%%
\subsection{Ks-band image}  
%%%%%%%%%%%%%%%%%%%%%%%%%%%%
\label{CFHTdata}
\textit{K}s-band imaging of NGC~6503 was obtained using WIRCAM on the CFHT in March 2009. The observations were carried out using a 16-point dither pattern repeated on each of the four WIRCAM chips, with 3$\times$15~s exposures at each position. This resulted in a total of $3\times16\times4=192~\times$ 15~s exposures, for a total integration time of 48 minutes. 

Processing of these data, including bad pixel rejection, sky subtraction and astrometric and photometric calibration, was performed using version 1.9 of the `I`iwi pipeline\footnote{\tt http://cfht.hawaii.edu/Instruments/Imaging/WIRCam/\\IiwiVersion1Doc.html}.  The photometric calibration, which is provided by the `I`iwi pipeline, was manually checked using background 2MASS stars in order to confirm its validity.  

\begin{figure}
\center
\includegraphics[scale=0.43]{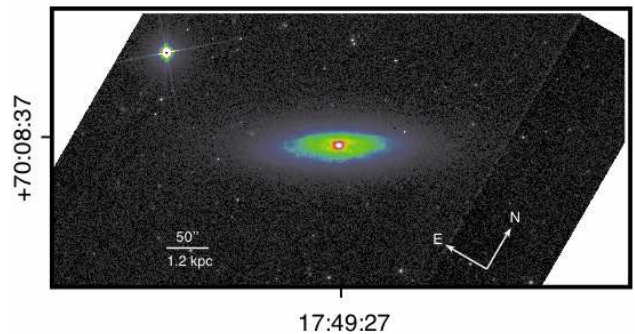}
\caption{CFHT WIRCAM $K$s-band image of NGC~6503, plotted on a logarithmic scale.   The extent of the image is the full length of the surface brightness profile discussed in Section~\ref{SBprofile}.  See Fig.~\ref{DBBmodel} for a sky-subtracted, masked view of the inner disc. (A colour version of this figure is available in the online journal.) \label{CFHTimg}}
\end{figure}

We estimated the sky level in the fully-reduced final image by calculating the mode of the sky intensities per pixel in 20 regions away from the galaxy that were free of any other sources.   The average and standard deviation of the sky values were also computed. The sky fluctuation is $\sim$~0.005\%, and we therefore subtracted a constant sky level across the entire image. The WIRCAM $K$s-band image of NGC~6503 is shown in Fig.~\ref{CFHTimg}.

%%%%%%%%%%%%%%%%%%%%%%%%%%%%
\section{\df}
%%%%%%%%%%%%%%%%%%%%%%%%%%%%
\label{discfit}
All of the model fits to the data presented in this paper were carried out using \df, a newly-released, publicly-available\footnote{\tt http://www.physics.rutgers.edu/$\sim$spekkens/diskfit/} code that fits non-parametric models to either images or velocity fields. While \df\ will fit axisymmetric models, its main purpose is to fit non-axisymmetric models either to images or to velocity fields of disc galaxies, as originally described by \citet{Reese07} for images and by both \citet{Spekkens07} and \citet{Sellwood10} for velocity fields. \df\ supercedes the \vf\footnote{\tt http://www.physics.rutgers.edu/$\sim$spekkens/velfit/} code to model velocity fields, and it represents the first public release of the non-parametric image decomposition technique described in \citet{Reese07}.  Both the photometric and kinematic branches of \df\ employ the same basic minimization technique, originally described by \citet{Barnes03}. Several new features have been added to the algorithm since the original publications: for example, \df\ can now fit for a symmetric outer velocity field warp, correct for minor distortions due to seeing, handle \FITS\ or text file velocity field inputs, and sparsely sample input \FITS\ files.

As described in detail in previous publications \citep{Reese07,Spekkens07,Sellwood10}, \df\ has several advantages over traditional approaches for extracting the kinematic or photometric structure of a disc galaxy.  The kinematic branch of the code, \vf, differs fundamentally from the frequently-used {\sc Rotcur}  \citep{Begeman89} ``tilted ring" algorithm since it fits a specific, physically motivated model rather than parametrizing concentric rings of the velocity field.  By the same token, the photometric branch of the code differs fundamentally from popular algorithms such as {\sc Galfit} \citep{Peng02, Peng10} in that it fits non-parametric disc and bar light profiles rather than specified functional forms.  \df\ therefore enables the user to model both kinematic and photometric data for nearby galaxies with as few assumptions about the parameters of the disc and bar as can reasonably be made. Because the underlying algorithm is the same for both the kinematic and photometric branches of the code, \df\ is a particularly powerful tool for understanding the physical structure of galaxies for which both types of data are available.

  \df\ excels at finding relatively weak, coherent asymmetries: it provides a robust estimate of the underlying disc properties in their presence, and quantitative upper limits in their absence. It is also superior to other algorithms in that it is capable of providing statistically valid, and realistic, estimates of the uncertainties on the returned parameters using bootstrap realizations of the best-fitting model to the data.
 
 Extensive documentation about the \df\ approach, models and capabilities can be found in previous publications  \citep{Reese07,Spekkens07,Sellwood10} and on the code website. Below, we summarize the salient features of the kinematic and photometric models that are relevant to the NGC~6503 analysis.

%%%%%%%%%%%%%%%%%%%%%%%%%%%%
\subsection{Kinematic Models}
%%%%%%%%%%%%%%%%%%%%%%%%%%%%
\label{vels}
The kinematic models implemented in \df\ are the same as in \vf, as described by \citet{Spekkens07} and \citet{Sellwood10}, with the additional capability of fitting a symmetric outer disc warp.  The current implementation of \df\ can fit the following model including rotation and non-circular flows of fixed azimuthal phase in a thin disc (c.f.\ eq.~5 in \citet{Spekkens07}):
\begin{eqnarray}
\nonumber
V_{model} & = & V_{sys} + \sin{i} \,\,\left[\, \bar V_t \cos{\theta}
\right. \\
&&  \left. - \; V_{m,t}\cos(m \theta_b) \cos{\theta}\, - 
\,V_{m,r}\sin( m \theta_b) \sin{\theta} \,\right] 
\label{bieq}
\end{eqnarray}
where $\bar V_t$ is the circular velocity, $V_{m,t}$ and $V_{m,r}$ are the tangential and radial components of non-circular flows with harmonic order $m=1$ or $m=2$ in the disc plane, $\theta$ and $\theta_b$ are the azimuthal angles relative to the major axis and the non-circular flow axis, respectively, and $i$ is the disc inclination. If $m=1$, the model describes a lopsided flow; if $m=2$ the model is bisymmetric, and describes a barred or elliptical flow.  \df\ therefore fits disc-plane models to the data, taking care of the projection of model components of order $m$ into sky-plane components of order $m^\prime = m \pm 1$. Unlike tilted-ring techniques \citep[e.g.][]{Schoenmakers97,Wong04}, the physical interpretation of the best-fitting \df\ models is therefore straightforward.

For consistency with other work \citep[e.g.][]{Simon05,Trachternach08}, the code can also fit radial flows, although such flows are physically poorly motivated.  This model assumes $m=0$ distortions to the flow in the disc plane, given by eq.~7 in \citet{Spekkens07}:
\begin{equation}
V_{model} = V_{sys} + \sin{i} \; \left[ \, \bar V_t \cos{\theta}
                         + V_r \sin{\theta} \, \right] \;,
\label{radeq}
\end{equation}
where $V_r$ is the radial flow component. 

\df\ requires a flat inner disc, but unlike its predecessor \vf, it allows for a symmetric warp in the outer disc.  The disc is assumed flat out to warp radius $r_w$, beyond which both the ellipticity and the position angle of the line of nodes vary in proportion to $(r-r_w)^2$.  \df\ can find the best fitting inner warp radius $r_w$ and peak change in ellipticity and position angle of the line of nodes, or it can hold any combination of these parameters fixed.

%%%%%%%%%%%%%%%%%%%%%%%%%%%%
\subsection{Photometric Models}
%%%%%%%%%%%%%%%%%%%%%%%%%%%%
\label{phot}
The photometric models applied by \df\ are identical to those described by \citet{Reese07}. \df\ fits photometric images with up to 3 components: a disc, bar and bulge. \df\ assumes a flat, intrinsically round disc and a linear bar with a different apparent (projected) ellipticity and position angle in the plane of the disc. Both the disc and bar light profiles are non-parametric, and thus do not have a specific functional form. 

The bulge is parametrized by the S\'ersic function
\begin{equation}
I(r) = I_0 \exp\left\{ -B_n \left[ \left( {r \over r_e} \right)^{1/n}
-1 \right] \right\},
\end{equation}
where $r_e$ is an effective radius, $n$ in the S\'ersic index, and $I_0$ is the intensity scale.  The constant $B_n$ is a function of the S\'ersic index, and is defined by the implicit relation $\Gamma(2n) = 2\gamma(2n,B_n)$ \citep{Graham01}. We allow the bulge to be spheroidal, with the disc plane being the plane of symmetry.

%%%%%%%%%%%%%%%%%%%%%%%%%%%%
\section{Results of Kinematic Modeling}
%%%%%%%%%%%%%%%%%%%%%%%%%%%%
\label{Kresults}
In this section, we present the best-fitting kinematic models to the H$\alpha$, CO and \HI\ velocity fields of NGC~6503.  For each of the velocity fields in Fig.~\ref{allvfs}, we use \df\ to fit for rotation only, as well as radial ($m=0$), lopsided ($m=1$) and bisymmetric ($m=2$) non-circular motions. We also search for a symmetric warp signature in the \HI\ velocity field.  We use \df\ to determine the galaxy centre, inclination, systemic velocity, and position angle of the kinematic major axis.  The minimum ring spacing is determined by the spatial resolution of the data and is increased to optimize the number of data points per ring, if necessary.   

We generate 1000 bootstrap realizations of each velocity field to determine uncertainties on the model parameters: we assume correlated residuals over 4-5\arcsec\ to derive uncertainties on the irregularly sampled H$\alpha$ data \citep[][the only available approach in \df\ for this data type]{Spekkens07}, and the radial/re-scaling method of \cite{Sellwood10} for the CO and \HI\ data.  The parameter $\Delta_{\mathrm{ISM}}$ is added in quadrature to the uncertainties in the emission line centroids during the fit \citep{Spekkens07}.  It is both an indicator of the turbulence in the disc and the relative size of the errors on the velocity field points.  We set the value of $\Delta_{\mathrm{ISM}}$ to produce reasonable $\chi^{2}$ statistics in the rotation-only model, and keep this parameter fixed for all other fits.  The relative differences in $\chi^{2}$ for each dataset thus indicate the goodness of fit.   We check that the model parameters and their uncertainties are unaffected by the particular value of $\Delta_{\mathrm{ISM}}$ or the bootstrap technique that is used. 

\begin{figure}
\center
\includegraphics[scale=0.41]{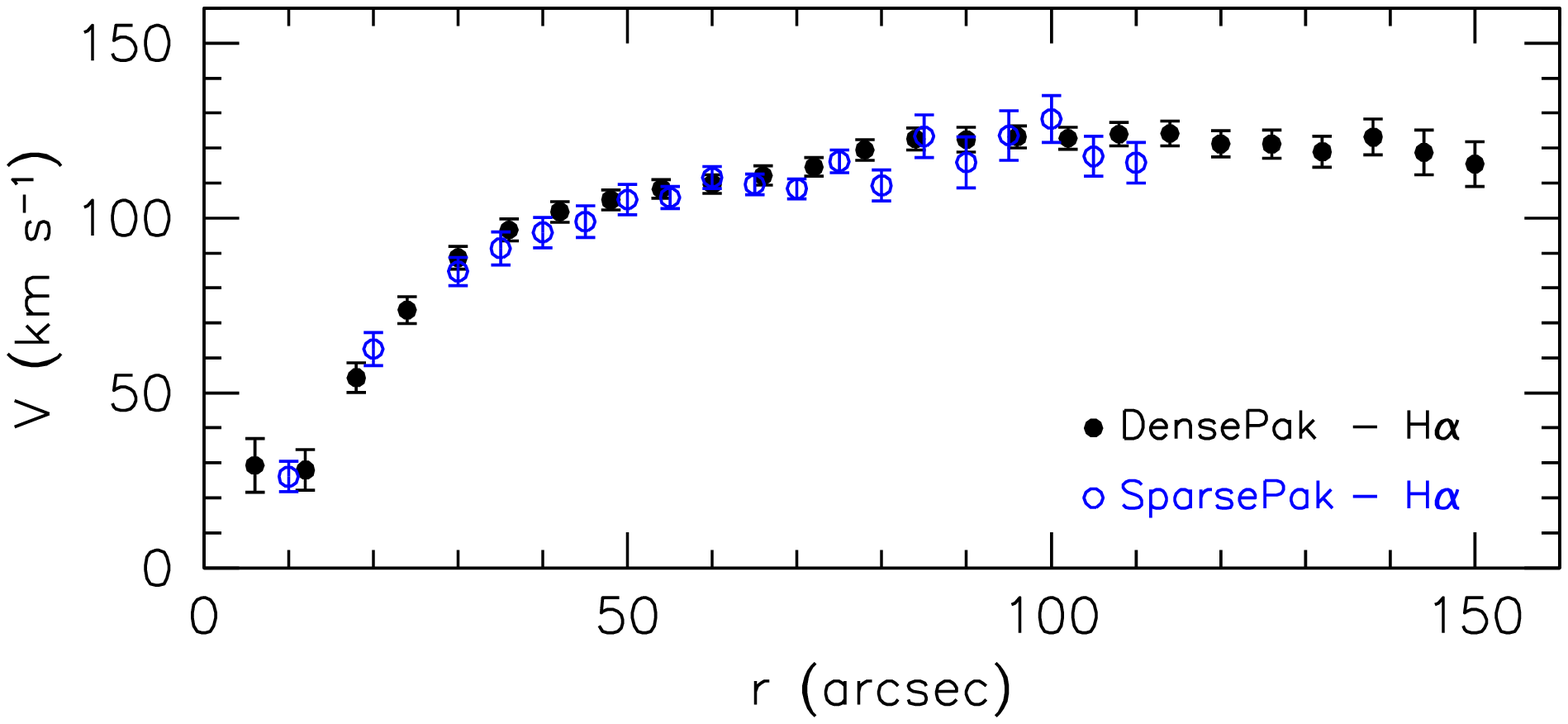}\\
\includegraphics[scale=0.41]{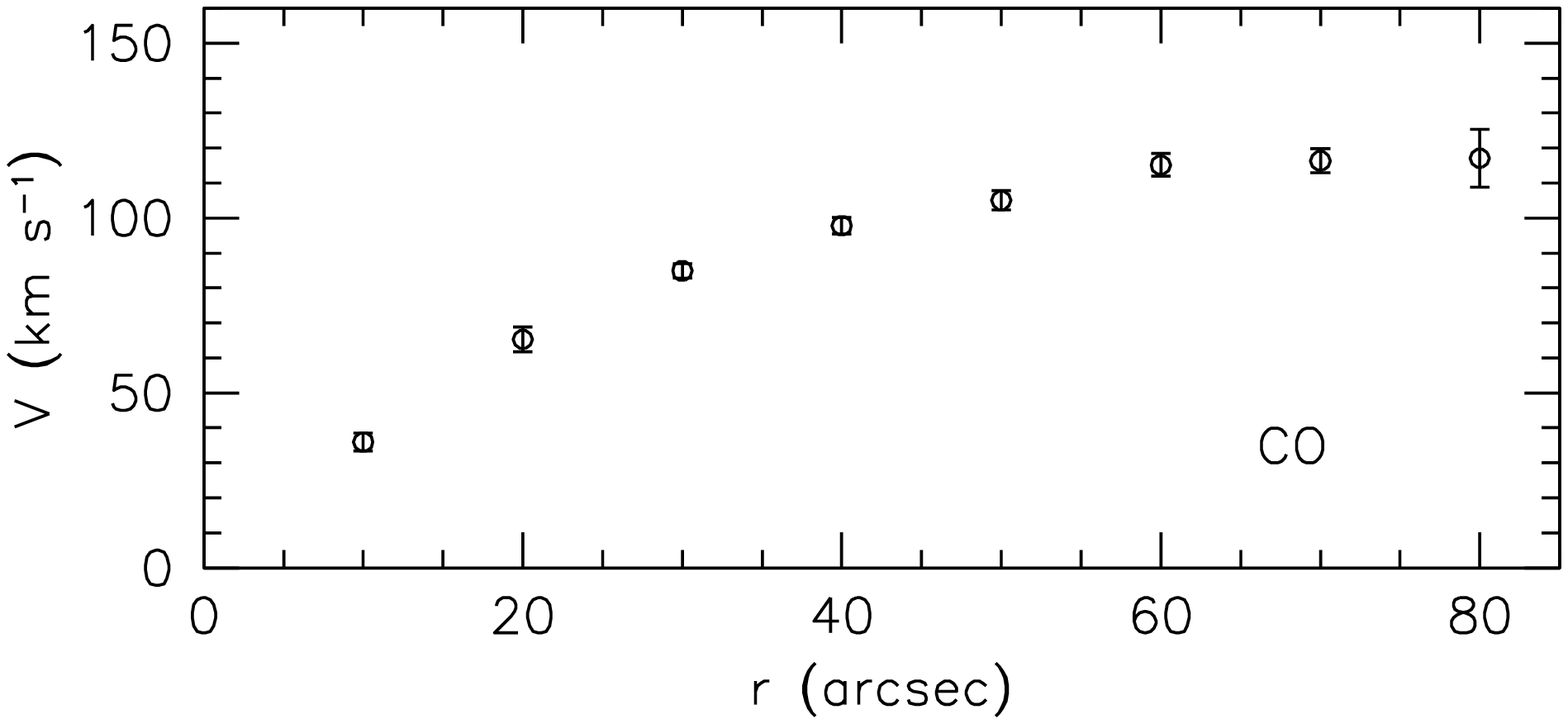}\\
\includegraphics[scale=0.41]{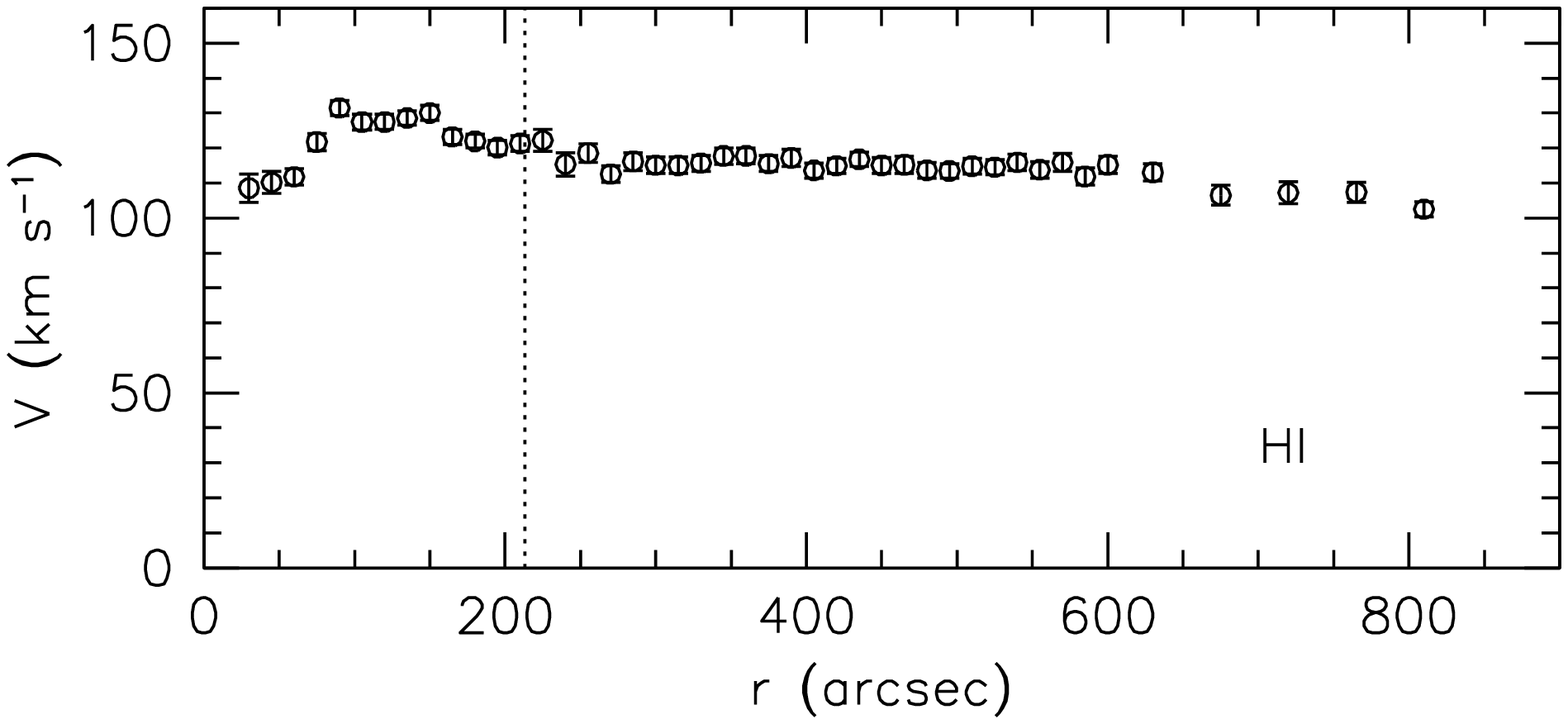}
\caption{Derived H$\alpha$ (top), CO (middle), and \HI\ (bottom) rotation curves for NGC~6503.  We find no evidence for coherent radial, lopsided, or bisymmetric non-circular motions in any of the datasets, nor a symmetric warp in the \HI\ data.  The vertical dotted line in the bottom panel indicates the edge of the optical disc \citep{deV91}. (A colour version of this figure is available in the online journal.) \label{allrcs}}
\end{figure}

We plot the derived rotation curves in Fig.~\ref{allrcs} and the \df\ rotation-only kinematic models and residuals in Fig.~\ref{rotmodelvfs}.   We also discuss the derived constraints on non-circular motions in the disc (Fig.~\ref{noncircs}).  We compare the results for each dataset (Fig.~\ref{allrcplot}) and to previous results in the literature.  Finally, we combine our multi-wavelength data and derive a master rotation curve for NGC~6503 (Fig.~\ref{masterrc}).  The best-fitting rotation-only model parameters for each dataset are listed in Table~\ref{datatable}.

%%%%%%%%%%%%%%%%%%%%%%%%%%%%
\subsection{H$\alpha$ rotation curves}
%%%%%%%%%%%%%%%%%%%%%%%%%%%%
\label{HArc}
We begin by fitting rotation-only models to both the DensePak and SparsePak velocity fields, and find that they provide a good representation of the data. The rotation velocities derived from the independent H$\alpha$ datasets agree well with each other, particularly for $10\arcsec \la r \la 80\arcsec$ (see Fig.~\ref{allrcs}).  The DensePak and SparsePak rotation curves both show a slow and smooth rise before flattening off at $V\sim$~120~\kms.  The derived values of the position angle (DPak: $-60.5\pm0.6\degr$; SPak: $-59.6\pm0.7\degr$) and systemic velocity (DPak: $28.4\pm1.0~\kms$; SPak: $28.4\pm1.7~\kms$) are identical to within their uncertainties.  The best-fitting inclinations (DPak: $71.0\pm0.9\degr$; SPak: $67.5\pm1.3\degr$) however, are different at about the 2$\sigma$ level.  Given that the radial extent of the SparsePak data is $\sim$1.5 times smaller than the DensePak data, these differences are not unexpected.  In addition, there are approximately 3.5 times more data points in the DensePak velocity field than the SparsePak velocity field.  That the adopted values of $\Delta_{\mathrm{ISM}}$ for the two H$\alpha$ velocity fields are different (6.5 \kms\ and 4.5 \kms\ for the DensePak and SparsePak data, respectively) is a reflection of the uncertainties on the individual fibre velocities being estimated in different ways for each dataset rather than an indication of a physical difference in the disc.

\begin{figure*}
\includegraphics[scale=0.36]{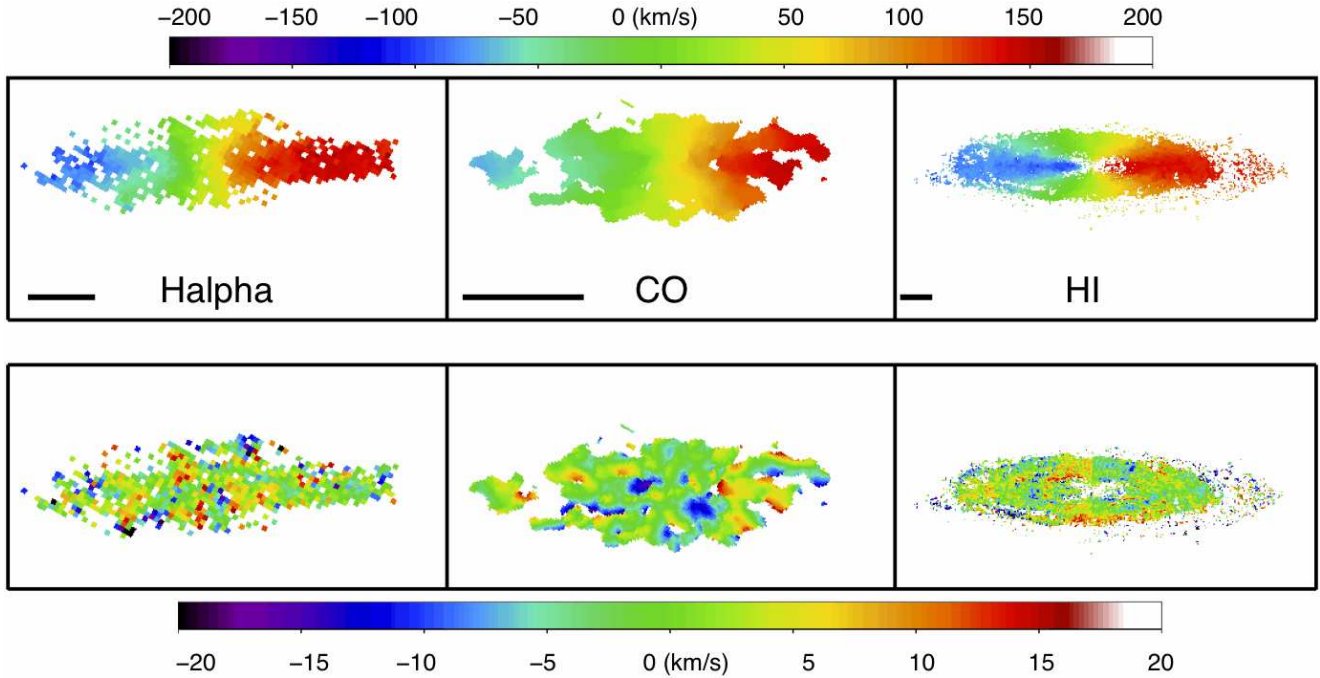}
\caption{Rotation-only \df\ models (top row) and residuals (bottom row) for the combined H$\alpha$ (left column), CO (middle column), and \HI\ (right column) data.  Note that the angular scale of each column is different: the horizontal bars in each panel of the top row are all 50$\arcsec$ (1.2~kpc) long.  See Fig.~\ref{allvfs} for the explicit dimensions of each velocity field. (A colour version of this figure is available in the online journal.) \label{rotmodelvfs}}
\end{figure*}

Because there is such broad agreement between the two derived rotation curves and sets of model parameters, we combine the DensePak and SparsePak velocity field data.  In Fig.~\ref{rotmodelvfs}, we show the combined H$\alpha$ \df\ rotation-only model and residuals. We also fit $m=(0,1,2)$ models to the combined H$\alpha$ velocity field; the tangential and radial components of these models are plotted in Fig.~\ref{noncircs}.  Radial, lopsided, and bisymmetric flows are consistent with 0~\kms\ throughout the optical disc, the change in $\chi^{2}$ when these flows are included is not statistically significant,  and the best-fitting disc parameters remain essentially unchanged from the rotation-only values in Table~\ref{datatable}. This suggests that the best physical model for the H$\alpha$ kinematics is one that contains rotation only. 

\citet{Epinat08} observed NGC~6503 as part of the GHASP survey.  Our final IFU H$\alpha$ rotation curve agrees well with the H$\alpha$ rotation curve they derived from their Fabry-P\'erot observations (see their figure E14). Both rotation curves extend to 150$\arcsec$ and reach similar maximum velocities (IFU: $126\pm3\,$\kms; Fabry-P\'erot: $117\pm9\,$\kms).  Their derived disc inclination ($72\pm2\degr$) and position angle ($-61\pm2\degr$) are also both in very good agreement with the values determined from our IFU data (see Table~\ref{datatable}).  

%%%%%%%%%%%%%%%%%%%%%%%%%%%%
\subsection{CO rotation curve}
\begin{figure}
\center
\includegraphics[scale=0.41]{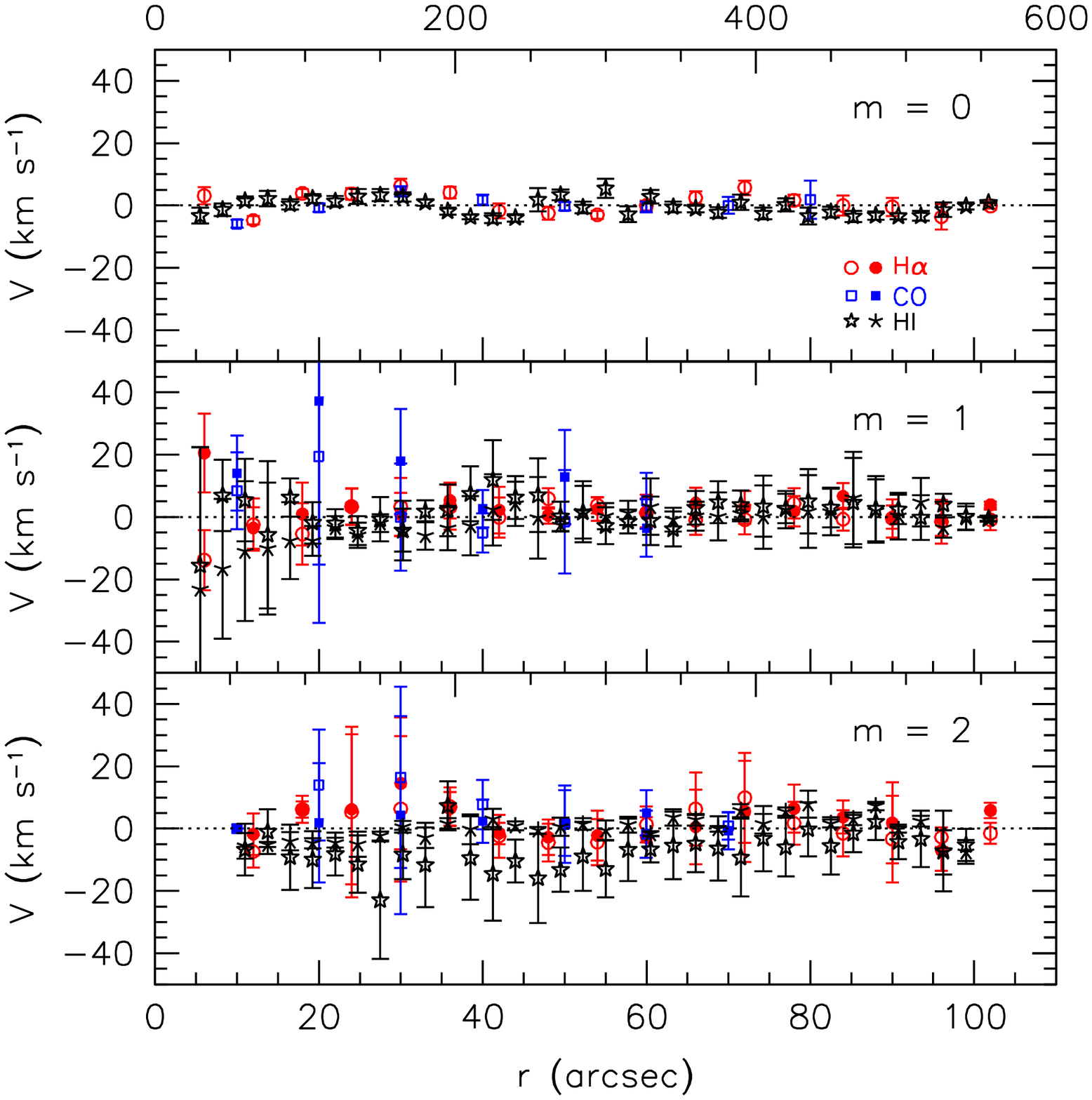}
\caption{Top: Radial ($m=0$) motions determined for the combined H$\alpha$ (circles), CO (squares), and \HI\ (stars) data. Middle: Tangential (open points) and radial (filled points) components of the lopsided ($m=1$) model. Bottom: Tangential (open points) and radial (filled points) components of the bisymmetric ($m=2$) model. (A colour version of this figure is available in the online journal.) \label{noncircs}}
\end{figure}

%%%%%%%%%%%%%%%%%%%%%%%%%%%%
\label{COrc}
In our best-fitting rotation-only model of the CO velocity field, the CO rotation curve (see Fig.~\ref{allrcs}) shows a smooth and steady rise in velocity ($V_{max}=117\pm8~\kms$) out to the last rotation curve point ($R_{max}=80\arcsec$).  The best-fitting disc parameters for this model are given in Table~\ref{datatable}. Fig.~\ref{noncircs} shows the tangential and radial components of $m=0,1,2$ models fit to the data. Similar to the H$\alpha$ data, the CO kinematics lack signs of organized radial, lopsided, or bisymmetric flows, and we therefore find the rotation-only model to be the best description of the CO velocity field (see Fig.~\ref{rotmodelvfs}).    

\citet*{Nishiyama01} derive a CO rotation curve using 16$\arcsec$ resolution observations from the 45-m Nobeyama Radio Observatory.  Their data show rising velocities out to a radius of $\sim55\arcsec$, followed by a region of constant velocity of $\sim125~\kms$ out to the end of the rotation curve just beyond 80$\arcsec$.  They determine the position angle of the major axis to be $-57\degr$, the inclination to be $70\degr$, and the maximum rotation velocity to be 127~\kms.  Our CO results for NGC~6503 are in good agreement with the \citet{Nishiyama01} data.

\begin{figure}
\center
\includegraphics[scale=0.41]{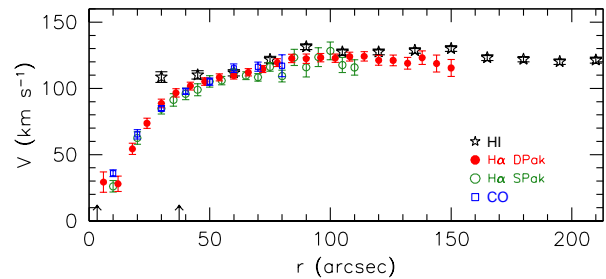}
\caption{Comparison of the derived rotation curves for NGC~6503 out to the edge of the optical disc.  The vertical arrows at 3.5$\arcsec$ and 37.5$\arcsec$ denote the positions of the circumnuclear disc and star-forming ring, respectively, that are discussed in Section~\ref{Features}.  (A colour version of this figure is available in the online journal.) \label{allrcplot}}
\end{figure}

\begin{figure}
\center
\includegraphics[scale=0.41]{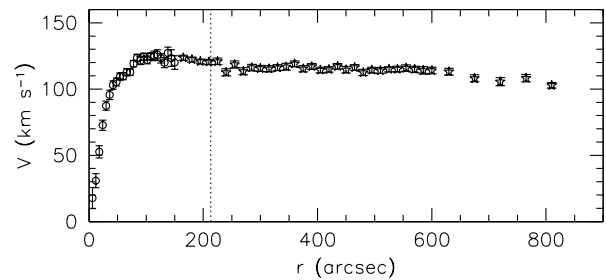}
\caption{Master rotation curve for NGC~6503.  The inner rotation curve is determined from the H$\alpha$ data (open circles) while the rotation at large radii is probed by \HI\ (stars).  The rotation curve data are given in Table~\ref{rcdata}.  The vertical dashed line shows the edge of the optical disc. \label{masterrc}}
\end{figure}

%%%%%%%%%%%%%%%%%%%%%%%%%%%%
\subsection{\HI\ rotation curve}
%%%%%%%%%%%%%%%%%%%%%%%%%%%%
\label{HIrc}
The \HI\ rotation curve (see Fig.~\ref{allrcs}) rises to a maximum velocity of $\sim$130~\kms\ around 100$\arcsec$ and then slowly declines out to $\sim$275$\arcsec$, just beyond the edge of the optical disc, before flattening off with a rotation speed of $\sim$115~\kms.  Although we recover the same rotation speed as reported by \citet{Greisen09}, those authors do not find a gradual decline at $100\arcsec \la r \la 200\arcsec$: this difference stems from our adoption of an envelope tracing model to estimate the velocity field from the \HI\ data cube rather than the first moment (Section~\ref{HIdata}).  Beyond 600$\arcsec$, the rotation curve shows hints of a very gradual decline, in rough agreement with the results of \citet{Greisen09}. 

We find radial, lopsided, and bisymmetric flows in the \HI\ data that are consistent with 0~\kms\ (see Fig.~\ref{noncircs}).  The \df\ velocity field model and residuals are shown in Fig.~\ref{rotmodelvfs}; like the optical and millimetre kinematic data, a rotation-only model provides the best description of the \HI\ velocity field.  The sharp line along the minor axis in the residual map is an artefact introduced during the Modified Envelope Tracing (see Section~\ref{HIdata}).  We also use \df\ to search for a symmetric warp in the \HI\ disc, and find no evidence for one: specifically, we find a 3$\sigma$ upper limit on the maximum change in position angle at $r \ga 30\arcsec$ to be ${\tt wphim} = 3\degr$, while that on the disc ellipticity is ${\tt welm =} 0.09$. The gradual decline in the \HI\ rotation curve for $r\ga 600\arcsec$ is therefore not caused by a warp. 

As in the models of \citet{Greisen09}, our best-fitting rotation-only \HI\ model underestimates the projected velocity in the outermost regions in the south/southwest quadrant (bottom right) of the disc by $10-15\,\kms$ (see Fig.~\ref{rotmodelvfs}).  The position-velocity diagrams of \citet{Greisen09} suggest that this feature is a local warp in the disc. That \df\ does not recover this feature as a warp is not surprising given its lack of symmetry and relatively small spatial extent.   

\begin{table*}
\begin{center}
\caption{Best-fitting kinematic parameters for each dataset.  Columns 2 and 3: position of centre relative to 17$^{\mathrm{h}}$49$^{\mathrm{m}}$26.4$^{\mathrm{s}}$ +70\degr08$\arcmin$40$\arcsec$.  Column 4: position angle of the kinematic major axis.  Column 5: inclination assuming the thin disc approximation.  The inclinations increase to a maximum of $\sim$~77$\degr$ if an intrinsic axis ratio of 0.2 is assumed.  Column 6: heliocentric systemic velocity.  Column 7: ISM gas turbulence adopted in the fit.  Column 8: Reduced chi-square of the best fitting model.  The last row of the table gives the parameters adopted in deriving the multi-wavelength master rotation curve presented in Section~\ref{Comparerc}.  \label{datatable}}
\begin{tabular}{|lccccccc|}
\hline
Dataset 			&$x_{c}$ 		&$y_{c}$ &P.A.  &\textit{i} 	&$V_{sys}$  &$\Delta$~ISM &$\chi^{2}_{r}$\\
 &($\arcsec$) &($\arcsec$) &(deg) &(deg) &(km s$^{-1}$) &(km s$^{-1}$) & \\
(1)  &(2) &(3) &(4) &(5) &(6) &(7) &(8)\\
\hline
H$\alpha$ - DPak 	&-0.9$\pm$0.7 	&0.1$\pm$0.5 	&-60.5$\pm$0.6 	&71.0$\pm$0.9 	&28.4$\pm$1.0 	&6.5 		&1.05\\
H$\alpha$ - SPak 	&-0.8$\pm$1.0 	&-1.1$\pm$0.8 	&-59.6$\pm$0.7 	&67.5$\pm$1.3 	&28.4$\pm$1.7 	&4.5 		&0.96\\
combined H$\alpha$ &-0.9$\pm$0.6 &0.1$\pm$0.5 	&-60.3$\pm$0.5 	&70.0$\pm$0.8 	&28.5$\pm$0.9 	&6.5 		&0.97\\
CO 				&0.3$\pm$1.1 	&-1.0$\pm$0.7 	&-60.5$\pm$0.5 	&73.7$\pm$1.1 	&31.4$\pm$2.5  	&2.5 		&0.99\\
\HI\ 				&1.1$\pm$0.3 	&-1.9$\pm$0.2 	&-60.2$\pm$0.3 	&73.5$\pm$0.4 	&24.7$\pm$0.7 	&6.0 		&0.98\\
\textbf{Master} 		&\textbf{-0.9} 	&\textbf{0.1} 	&\textbf{-60.2} 		&\textbf{73.5} 		&\textbf{24.7} 		&... 		&...\\
\hline
\end{tabular}
\end{center}
\end{table*}

%%%%%%%%%%%%%%%%%%%%%%%%%%%%
\subsection{Comparison of velocity data}
%%%%%%%%%%%%%%%%%%%%%%%%%%%%
\label{Comparerc}
The rotation curves for NGC~6503 derived using different tracers are all well-behaved and show no signs of radial ($m=0$), lopsided ($m=1$) or bisymmetric ($m=2$) non-circular motions.  In Figure \ref{allrcplot}, we directly compare the H$\alpha$, CO, and \HI\ rotation curves.  The rising part of the rotation curve is well-mapped by the CO and H$\alpha$ data.  The overlap between the CO and H$\alpha$ rotation velocities suggests that the optical data out to $\sim$80$\arcsec$ are not significantly affected by extinction \citep[e.g.][]{Bosma92}.  Beyond $\sim$45$\arcsec$, the \HI\ rotation curve agrees well with the H$\alpha$ and CO data; the discrepant \HI\ rotation curve point at $r\sim30\arcsec$ is unreliable because it is likely  biased by beam smearing.

The position angle of the kinematic major axis is well-constrained, with all three datasets providing nearly indistinguishable results (combined H$\alpha$: $-60.3\pm0.5\degr$; CO: $-60.5\pm0.5\degr$; \HI: $-60.2\pm0.3\degr$).  The disc inclinations derived from the CO ($73.7\pm1.1\degr$) and \HI\  ($73.5\pm0.4\degr$) are in good agreement, suggesting the inner and outer discs are coplanar.  This is consistent with \df\ not finding a warp in the \HI\ data (see Section~\ref{HIrc}).  The H$\alpha$-derived inclination ($70.0\pm0.8\degr$), however, is lower by about 2.5$\sigma$.  There is a similar difference between the \HI\ ($24.7\pm0.7~\kms$) and H$\alpha$ ($28.5\pm0.9~\kms$) systemic velocities.   

The agreement between the best-fitting model parameters and similarities between the rotation curves for each dataset suggest that all are tracing the same underlying disc kinematics.  For these reasons, we determine a master rotation curve for NGC~6503 by combining the higher-resolution H$\alpha$ data across the inner disc and the \HI\ data farther out; we use the H$\alpha$-derived galaxy centre and the \HI-derived inclination, position angle, and systemic velocity.  The master rotation curve is shown in Figure \ref{masterrc}, and the adopted disc geometry is given in Table~\ref{datatable}.  The rotation curve data are given in Table~\ref{rcdata}.

%%%%%%%%%%%%%%%%%%%%%%%%%%%%
\section{Results of Photometric Modeling}
%%%%%%%%%%%%%%%%%%%%%%%%%%%%
\label{Presults}
In this section, we present the best-fitting photometric models to the $K$s-band image of NGC~6503.  We use \df\ to fit disc-only, disc+bulge, disc+bar, and disc+bar+bulge models.  \df\ determines the galaxy centre, disc inclination and position angle, the S\'ersic index, effective radius and ellipticity of the bulge, and the bar position angle and ellipticity.  In addition, \df\ calculates the percentage of light coming from each component (disc, bulge, bar) of the model.  We generate 1000 bootstrap realizations of the image to determine uncertainties on the model parameters. 

We plot the best-fitting \df\ models and residuals in Figures~\ref{DBBmodel}, \ref{badphotfits} and \ref{componentsfig}; the values of the model parameters are listed in Table~\ref{phottable}.  We discuss the photometric signatures of morphological features in the disc (Fig.~\ref{structurefig}).  Finally, we present the derived $K$s-band surface brightness profile for NGC~6503 in Figures~\ref{sbprofile} and \ref{sbfit} and compare it to previous results in the literature.

%%%%%%%%%%%%%%%%%%%%%%%%%%%%
\subsection{\df\ Photometric Models}
%%%%%%%%%%%%%%%%%%%%%%%%%%%%
\label{discfitmodels}

\begin{figure}
\center
\includegraphics[scale=0.95]{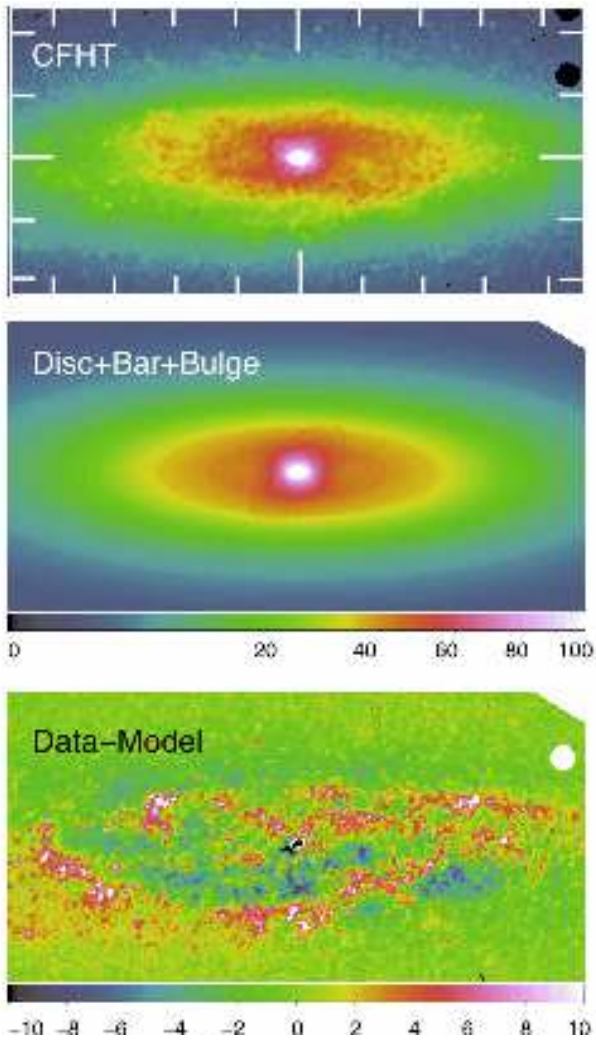}
\caption{Masked, sky-subtracted $K$s-band image of NGC~6503 (top), the best-fitting \df\ disc+bar+bulge model (middle), and the residuals (bottom).  Tickmarks in the top panel are spaced every 15$\arcsec$ and the image has been rotated 30$\degr$ to the west.  The full radial extent of the CFHT image is $\sim$~370$\arcsec$; the panels are cropped to exclude most of the sky.   The units on the colourbars are ADU. (A colour version of this figure is available in the online journal.) \label{DBBmodel}}
\end{figure}

\begin{figure}
\center
\includegraphics[scale=0.60]{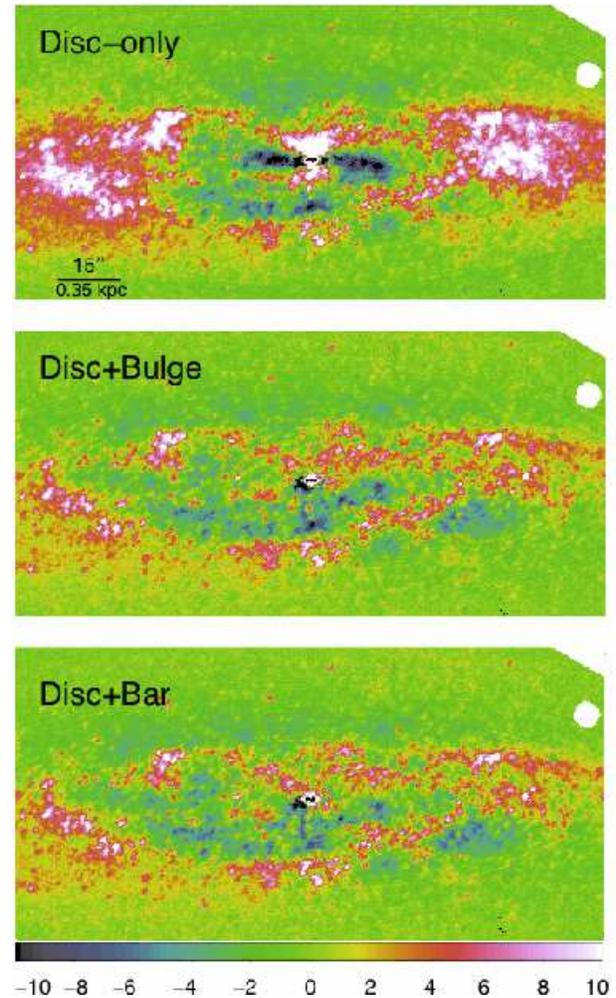}
\caption{Residuals for the disc-only (top), disc+bulge (middle) and disc+bar (bottom) models.  The size of each panel is the same as in Fig.~\ref{DBBmodel}; the horizontal bar in the lower left corner of the top panel is 15$\arcsec$ long and the models have been rotated 30$\degr$ to the west.  The units on the colourbar are ADU. (A colour version of this figure is available in the online journal.) \label{badphotfits}}
\end{figure}

\begin{figure}
\center
\includegraphics[scale=0.50]{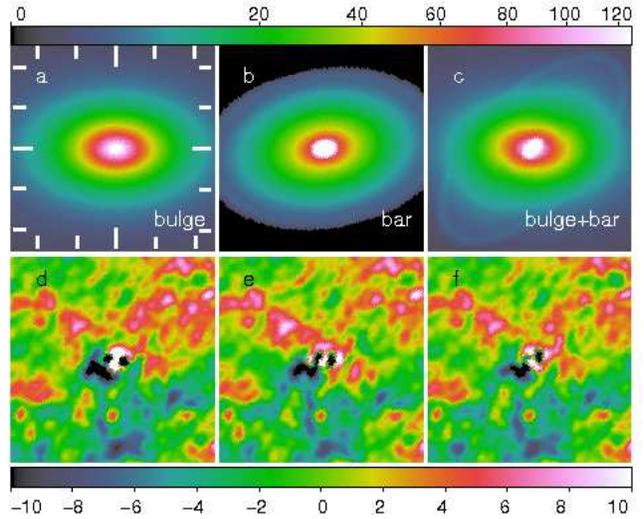}
\caption{(a): bulge component of the disc+bulge model.  (b): bar component of the disc+bar model.  (c): bar and bulge components of the disc+bar+bulge model.  The bar is the structure stretching from the lower left corner to the upper right.  (d): residuals of the disc+bulge model.  (e): residuals of the disc+bar model.  (f): residuals of the disc+bar+bulge model.  Each panel is 25$\arcsec$$\times$25$\arcsec$ in size and the models have been rotated 30$\degr$ to the west so that the disc major axis is horizontal.  The tickmarks in panel a are spaced every 5$\arcsec$.  The units on the upper and lower colourbars are ADU.  In the two-component models (panels a and b), the non-disc structures are very similar except for their position angles relative to the disc major axis.  In the three-component model (panel c), the bar is more elliptical and angled farther away from the disc major axis than the bar in the disc+bar model (panel b). \label{componentsfig}}
\end{figure}

With different combinations of disc, bulge and bar components, we fit four \df\ models to the CFHT data (see Figures~\ref{DBBmodel}, \ref{badphotfits} and \ref{componentsfig}).  The derived galaxy centre and the position angle of the disc are identical to within their uncertainties for all four models (see Table~\ref{phottable}).  These disc position angles are also in very good agreement with the kinematically-derived position angle (see Table~\ref{datatable}).  With the exception of the disc-only model, the derived disc inclinations are, within their errors, consistent with each other and the kinematically-derived inclination.  

The disc-only model is a poor fit to the data.  The derived disc inclination is comparatively low, and the residuals are large in the outer parts of the disc, as well as near the centre (see Fig.~\ref{badphotfits}).  The low inclination is an indication that there is a relatively bright circular component/feature at the galaxy centre that the model is trying to fit.  In its innermost regions, NGC~6503 is more than a pure disc galaxy.  

The two-component (disc+bulge and disc+bar) models are an improvement over the disc-only model.  In both of these cases, the disc contributes the majority of the light ($\sim$~95\%) and the models do a good, and similar, job of fitting the outer disc (see Fig.~\ref{badphotfits}). The smaller residuals in Fig.~\ref{badphotfits} show this qualitatively, as do the smaller $\chi^{2}_{r}$ values for the disc+bulge and disc+bar models in Table~\ref{phottable}; the improvement in the fit is highly statistically significant, even accounting for the extra parameters in the fit.

With all of the bulge parameters ($r_{e}$, $n$, $\epsilon$) in the disc+bulge model allowed to vary, the S\'ersic index of the bulge is poorly constrained.  We therefore run two models and fix the bulge S\'ersic index to $n$~=~1.1 and $n$~=~1.9.  We choose these values based on the results of \citet{Puglielli10} who model the $B$- and $R$-band surface brightness profiles of \citet{Bottema89}.  They find the galaxy bulge to be nearly exponential ($n$~=~1.1, $r_{e}$~=~6.4$\arcsec$) or to be in the range of a pseudobulge ($n$~=~1.9, $r_{e}$~=~14$\arcsec$), depending on whether they model the entire surface brightness profile, or exclude the Type II hump (see the discussion in Section~\ref{SBprofile} below).  

The results for the $n$~=~1.1 and $n$~=~1.9 \df\ models are both qualitatively and quantitatively very similar, and the \df-derived effective radius of the bulge when $n$~=~1.1 is the same as that determined by \citet{Puglielli10} to within the errors.  We list the values of the $n$~=~1.1 model parameters in Table~\ref{phottable} and note here that, to within the errors, the disc parameters of the $n$~=~1.9 model are the same as those derived in the $n$~=~1.1 model and that the $n$~=~1.9 bulge is slightly larger ($r_{e}$~=~8.9$\arcsec$) and more round ($\epsilon$~=~0.34) than the $n$~=~1.1 bulge.  For concreteness, we adopt the $n$~=~1.1 bulge model parameters in what follows, although the analysis remains unchanged if the $n$~=~1.9 results are used.

In Fig.~\ref{componentsfig}a we show the $n$~=~1.1 \df\ bulge. In the disc+bulge model, the bulge is not spherical, but rather has a projected ellipticity along the disc major axis of $\epsilon=0.4$.  It contains 5.4\% of the total light and has an effective radius of  $r_{e}$~=~6.7$\arcsec$.

The bar determined by \df\ in the disc+bar model is relatively round ($\epsilon$~=~0.39) and oriented $\sim$~7$\degr$ away from the disc major axis in projection. It contains 4.5\% of the total light.  The bar is shown in Fig.~\ref{componentsfig}b. 

The bar model does a better job of fitting the data than the bulge model ($\chi^{2}_{r, bar}$~=~2.72 vs.~$\chi^{2}_{r, bulge}~=~3.05$).  This can also be seen by looking closely at the residuals in Fig.~\ref{componentsfig}d and Fig.~\ref{componentsfig}e.  In particular, in the bulge model, the large $\pm$10~ADU residuals at the centre, the red/pink regions of 5-8~ADU residuals in the upper right quadrant and the extended red region of 5~ADU residuals at the left side of the panel of Fig.~\ref{componentsfig}d are greatly improved in the bar model (Fig.~\ref{componentsfig}e).  The trade-off in the bar model are larger residuals 90$\degr$ away from the bar in the upper left and lower right quadrants.

The similarity between the morphology of the bulge in the disc+bulge model and that of the bar in the disc+bar model (Fig.~\ref{componentsfig}a and Fig.~\ref{componentsfig}b) illustrates that the bar component is trying to encompass the bulge of the galaxy, while simultaneously trying to fit an asymmetric component that is not aligned with the major axis of the disc.  The non-disc components of the disc+bulge model and disc+bar model are therefore both attempting to fit the same feature in the light distribution (Fig.~\ref{componentsfig}).  The primary difference between the models is whether or not the asymmetry is allowed to move away from the disc major axis.

We are able to improve the fit in the entire central region of the galaxy by allowing the \df\ model to contain both a bar and a bulge (see Fig.~\ref{DBBmodel} and Fig.~\ref{componentsfig}).  In this three-component model, the disc contains 94.7\% of the total light and its geometry agrees well with those derived from the kinematic data.  The bar contains  0.7\% of the light, has an ellipticity of $\epsilon=0.53$, and projects near the minor axis of the disc (PA$_{bar}\sim-17\degr$).  As the bulge parameters are unconstrained if they are all allowed to vary, we fix the values to those found in the $n$~=~1.1 disc+bulge model.  Changing the bulge properties does not make a difference to the \df\ model: it is the addition of an extra physical component that matters.  The bulge contains 4.6\% of the total light.

The modest improvement in $\chi^{2}_{r}$ for the three-component disc+bar+bulge model (see Table~\ref{phottable}) is statistically significant at the 4$\sigma$ level due to the large number ($>$~37500) of degrees of freedom.  Moreover, we consider the three-component model to be the best physical description of NGC~6503 because the residuals in Fig.~\ref{componentsfig}  highlight the need for both a component in the disc and an extended asymmetric structure that is not aligned with the disc major axis.

%%%%%%%%%%%%%%%%%%%%%%%%%%%%
\subsection{Photometric Signatures of Morphological Features}
%%%%%%%%%%%%%%%%%%%%%%%%%%%%
\label{Features}

\begin{figure}
\center
\includegraphics[scale=0.65]{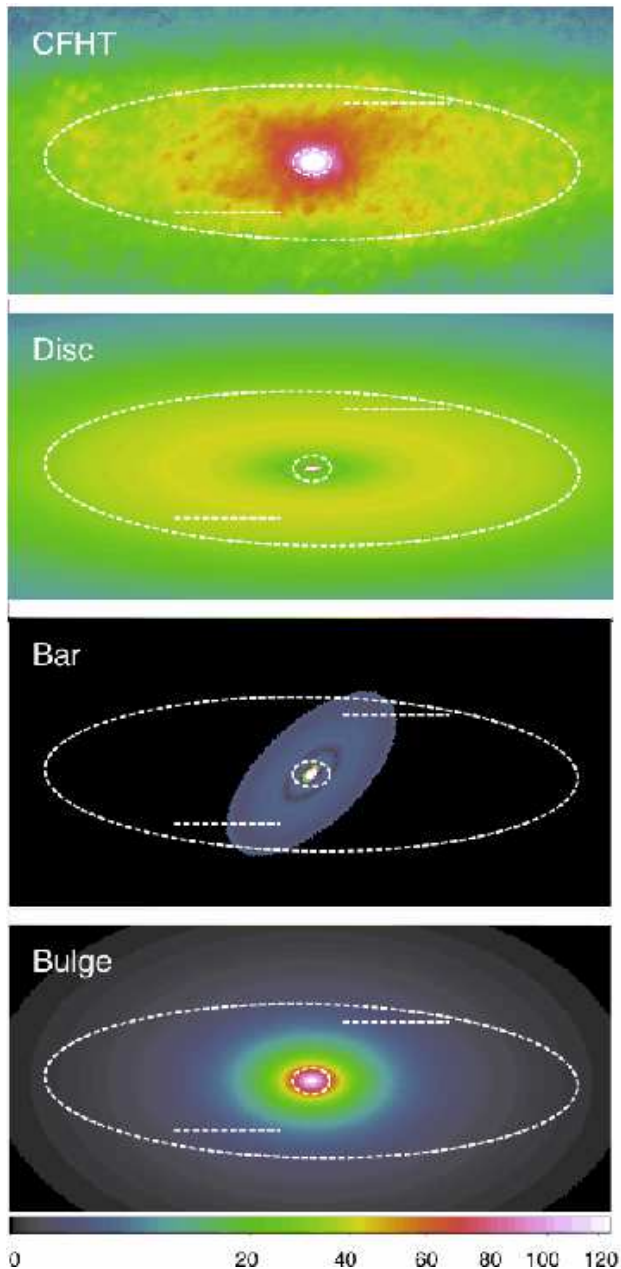}
\caption{Observed $K$s-band image of NGC~6503 (top) and the individual components of the \df\ disc+bar+bulge model.  The large and small dashed ellipses indicate the positions of the star-forming ring and circumnuclear disc, respectively, that are discussed by \citetalias{Freeland10}.  To guide the eye, the dashed lines are parallel to the spiral arms coming off the ends of the end-on bar; for comparison to Fig.~\ref{DBBmodel}, these lines are 15$\arcsec$ (0.35~kpc) in length.  The image has been rotated 30$\degr$ to the west. (A colour version of this figure is available in the online journal.) \label{structurefig}}
\end{figure}

Using a multi-wavelength dataset, \citetalias{Freeland10} determine that the disc of NGC~6503 contains several structures.  They find that there is a 3.5$\arcsec$ circumnuclear disc that is closely aligned to the major axis of the galaxy disc and that a nuclear spiral exists within the circumnuclear disc.  They also find a 7$\arcsec$ end-on bar; the galaxy's spiral arms extend from the bar ends.  Also present is a 37.5$\arcsec$ star-forming ring.  

In Fig.~\ref{structurefig}, we indicate the positions of the features highlighted by \citetalias{Freeland10} in relation to the disc, bar and bulge components determined by \df\ in the three-component model.  The \df\ bulge is clearly fitting the circumnuclear disc.  \df\ finds a relatively bright bar-like component within the circumnuclear disc, although the negative residuals in this region (see Fig.~\ref{componentsfig}f) suggest this to be an artefact of the fit. On the other hand, the faint bar-like component beyond the circumnuclear disc encloses an elliptical area that roughly connects the galaxy's spiral arms: this recovered structure is likely real.  We note that Fig.~\ref{structurefig} has been rotated 30$\degr$ to the west making the close-to-end-on bar (PA$_{bar}\sim-17\degr$) appear closer to $\sim-45\degr$ in that figure.   The star-forming ring is not prominent in the \df\ model nor the $K$s-band image, which is not surprising given that the $K$s-band probes the old stellar population.

The $K$s-band photometric modeling with \df\ shows NGC~6503 to be a disc-dominated system that has small, but measurable, non-disc structure.  To summarize, the disc+bar+bulge model described in Section~\ref{discfitmodels} is not only the best statistical description of the data, but also the best physical model when compared to structures found by \citetalias{Freeland10}.  To accurately model the inner region of the galaxy both on and off the disc major axis, two separate non-disc components are required.  In addition, the bar that \df\ finds is positioned very similarly to the bar found by \citetalias{Freeland10}.

\begin{table*}
\begin{center}
\caption{Best-fitting photometric parameters. Rows 1 and 2: position of centre relative to 17$^{\mathrm{h}}$49$^{\mathrm{m}}$26.4$^{\mathrm{s}}$ +70\degr08$\arcmin$40$\arcsec$.  Row 3: position angle of the disc major axis.  Row 4: inclination of the disc assuming the thin disc approximation.  The inclinations increase to a maximum of $\sim$~77$\degr$ if an intrinsic axis ratio of 0.2 is assumed.  Row 5: effective radius of the bulge.  Row 6: Adopted S\'ersic index of the bulge.  Row 7: ellipticity of the bulge.  Row 8: position angle of the bar in the sky plane.  Row 9: ellipticity, defined as 1-(b/a), of the bar.    Row 10: reduced chi-square of the model.  Row 11: fraction of light attributed to the disc.  Row 12: fraction of light attributed to the bulge.  Row 13: fraction of light attributed to the bar.    \label{phottable}}
\begin{tabular}{|lcccc|}
\hline
  									&Disc-only		&Disc+Bulge  		&Disc+Bar  		&Disc+Bar+Bulge\\ 
\hline
(1) x$_{c}$ ($\arcsec$)		&0.70$\pm$0.05	&0.70$\pm$0.06 	&0.70$\pm$0.04 	&0.70$\pm$0.02\\
(2) y$_{c}$ ($\arcsec$)		&-0.50$\pm$0.03 	&-0.50$\pm$0.03	&-0.50$\pm$0.02 	&-0.50$\pm$0.02\\
(3) Disc P.A. (deg)			&-60.0$\pm$2.6 	&-60.1$\pm$1.0	&-60.3$\pm$0.7 	&-60.3$\pm$0.7\\
(4) Disc inc (deg)			&68.7$\pm$2.5 	&72.3$\pm$1.9		&72.4$\pm$0.9 	&72.9$\pm$0.8\\
(5) Bulge $r_{e}$ ($\arcsec$)	&... 				&6.7$\pm$0.7		&... 				&6.7\\
(6) $n$					&... 				&1.1				&... 				&1.1\\
(7) Bulge $\epsilon$			&... 				&0.41$\pm$0.05	&... 				&0.41\\
(8) Bar P.A. (deg)			&... 				&...				&-52.8$\pm$4.4 	&-16.7$\pm$8.5\\
(9) Bar $\epsilon$			&... 				&...				&0.39$\pm$0.05 	&0.53$\pm$0.04\\
(10) $\chi^{2}_{r}$			&4.72 			&3.05			&2.72 			&2.53\\
(11) \% Disc		 		&100 			&94.6$\pm$0.9		&95.5$\pm$0.5 	&94.7$\pm$0.5 \\
(12) \% Bulge 				&... 				&5.4$\pm$0.9		&... 				&4.6$\pm$0.5\\
(13) \% Bar 				&... 				&...				&4.5$\pm$0.5 		&0.7$\pm$0.3 \\

\hline
\end{tabular}
\end{center}
\end{table*}

\begin{figure}
\center
\includegraphics[scale=0.41]{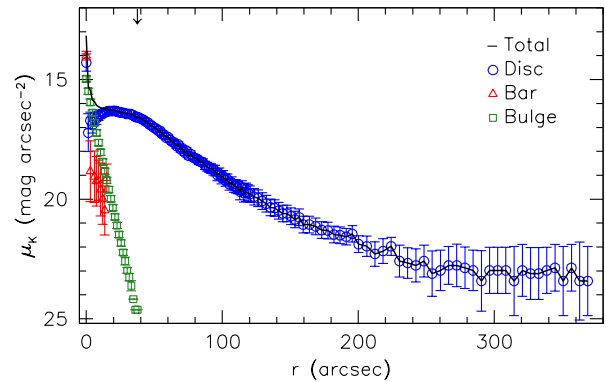}
\caption{Surface brightness profile for NGC~6503 derived from the \df\ disc+bar+bulge model.  The solid black line is the total light.  The blue open circles are the contribution from the disc.  The red open triangles are the contribution from the bar.  The green stars are the contribution from the bulge.  The arrow at 37.5$\arcsec$ indicates the position of the star-forming ring.  (A colour version of this figure is available in the online journal.) \label{sbprofile}}
\end{figure}

\begin{figure}
\center
\includegraphics[scale=0.41]{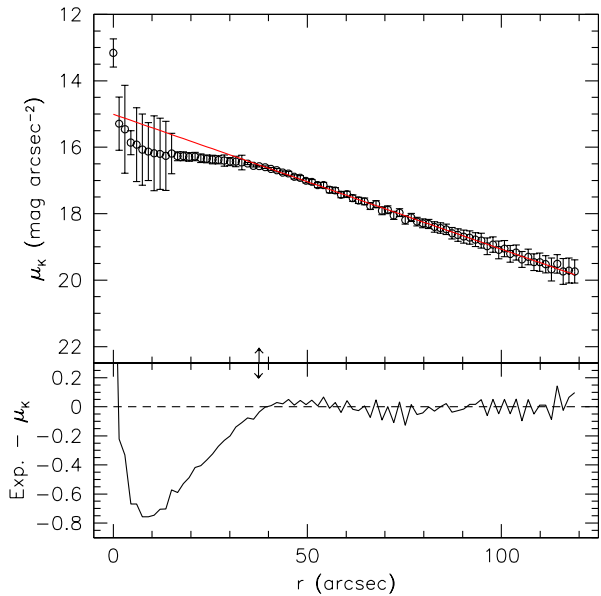}
\caption{Top panel: total inner \df\ surface brightness profile for NGC~6503.  The red line is an exponential fit to the data between 40$\arcsec$ and 120$\arcsec$.  Bottom panel: difference between the exponential fit and the observed surface brightness profile.  The vertical arrows indicate the position of the star-forming ring from \citetalias{Freeland10} that coincides with the Type II plateau in the surface brightness profile.  (A colour version of this figure is available in the online journal.) \label{sbfit}}
\end{figure}

%%%%%%%%%%%%%%%%%%%%%%%%%%%%
\subsection{Surface Brightness Profile}
%%%%%%%%%%%%%%%%%%%%%%%%%%%%
\label{SBprofile}
In Fig.~\ref{sbprofile}, we show the $K$s-band surface brightness profiles derived from this disc+bar+bulge model.  Note that \df\ automatically produces separate profiles for each of the model components.  The high central point of the bar profile is unphysical and is likely being confused with the bulge component.  This means that the fraction of the total light that is assigned to the bar (0.7\%) is actually an over-estimate.

The radial extent of the surface brightness profile in Fig.~\ref{sbprofile} is three times farther than previous data compiled by \citetalias{Freeland10}.  The profile displays a Freeman Type II behaviour \citep{Freeman70}, being exponential at large radii, having a plateau between 10$\arcsec$ and 40$\arcsec$, and rising steeply inside of 10$\arcsec$.  To highlight the plateau, in Fig.~\ref{sbfit} we fit an exponential to the profile at radii between $\sim$~40$\arcsec$ and 120$\arcsec$ and plot the difference between the exponential fit and the observed profile.  There is a maximum difference between the two of about 0.8 mag arcsec$^{-2}$ at a radius of $\sim$~10$\arcsec$.  The position of the star-forming ring discussed by \citetalias{Freeland10} coincides with the outer radius of the plateau in the surface brightness profile.  The outer profile beyond 120$\arcsec$ (Fig.~\ref{sbprofile}) is not exponential.  We have verified that this is a real feature of the galaxy and not an artefact related to sky subtraction \citep[e.g.][]{McDonald09}.  

\citetalias{Freeland10} present an $H$-band surface brightness profile of NGC~6503, along with $B$- and $R$-band profiles from \citet{Bottema89}, and 3.6$\mu$m and 4.5$\mu$m profiles from the \textit{Spitzer} archive.  These profiles show the same Type II behaviour at the same radii as our $K$s-band profile.  In addition, the magnitude of the average deviation of these data from an exponential profile is consistent with what we find for the $K$s-band data.

%%%%%%%%%%%%%%%%%%%%%%%%%%%%
\section{Does NGC~6503 Contain a Strong Bar?}
%%%%%%%%%%%%%%%%%%%%%%%%%%%%
\label{Bar}

\citet{Bottema89} argues that dust extinction is the cause of the plateau observed in the $B$- and $R$-band surface brightness profiles of NGC~6503.  We can rule out this explanation as not only is the Type II profile observed in our infrared $K$s-band data, but the agreement between our derived H$\alpha$ and CO rotation curves also suggests minimal extinction.  Based on simulations by \citetalias{Bureau05}, \citetalias{Freeland10} argue that a better explanation for the Type II surface brightness profile in NGC~6503 is the presence of a bar.  \citet{Puglielli10} similarly prefer the bar explanation based on their modeling and results of \citet{Foyle08} and \citet{Bournaud05}. 

\defcitealias{Emsellem01}{Emsellem et al.~2001}
\defcitealias{Rix92}{Rix et al.~1992}

\citetalias{Freeland10} provide a number of lines of evidence to support the idea of a bar.  Nuclear spirals and star-forming rings are resonance features of bars \citep[e.g.][]{Athanassoula99}, and both are observed in NGC~6503.  There is a significant central drop ($\sigma$-drop) in the stellar velocity dispersion of the galaxy \citep[e.g.][]{Bottema89}.  $\sigma$-drops can be caused by a number of mechanisms including bars, dynamically cold gas forming stars in a circumnuclear disc, or even counter-rotating stellar discs \citepalias[e.g.][]{Bureau05, Emsellem01, Rix92}.  While \citetalias{Freeland10} find evidence for a circumnuclear disc in their WHIRC $H$-band imaging data, they argue that the $\sigma$-drop is more likely caused by a bar, given the presence of the nuclear spiral and star-forming ring.  

\citetalias{Freeland10} specifically indicate that it is a \textit{strong} end-on bar.  \citetalias{Bureau05} run $N$-body simulations to highlight how the bar strength and orientation (with respect to the observer's line-of-sight) and galaxy inclination affect, among other things, the shape of the $\sigma$ profile.  In these simulations, only a strong end-on bar can cause a central $\sigma$-drop.  As in \citetalias{Freeland10}, we adopt this qualitative distinction between a strong bar and a weak one for the following discussion.  

Do we find any \textit{direct} evidence for a strong bar in our dataset and \df\ models?  Strong bars should produce disturbed kinematics similar to those seen in figure 1 of \citetalias{Bureau05}.  Our H$\alpha$, CO, and \HI\ velocity fields and derived rotation curves are well-described by rotation-only \df\ models and agree remarkably well with each other.  With the exception of a possible hint of the broadening of the velocity profile in the central pixel of the CO data, we do not detect the bar in the rotation curves or velocity fields.   

Though we can see hints of the star-forming ring in the H$\alpha$ and CO intensity maps (see Figures~\ref{dpharing} and \ref{COmom0}), we do not detect it kinematically (see Figures~\ref{rotmodelvfs} and \ref{allrcplot}), suggesting that the bar is not currently strong enough to be funneling gas.   Additionally, there is no build-up of CO along the ring near the ends of the bar (see Fig.~\ref{COmom0}), which one would expect if the bar were strong \citep[e.g.][]{Lin08}. We estimate the molecular gas mass at 30$\arcsec~<~r~<$~40$\arcsec$ $-$ the location of the star-forming ring $-$ to be 4.2~$\times$~10$^{6}$~M$_{\odot}$ (using X$_{\mathrm{CO}}$~=~2~$\times$~10$^{20}$ cm$^{-2}$/K \kms\ and $\alpha_{\mathrm{CO}}$~=~4.3~M$_{\odot}$/K \kms; \citealt{Solomon05}), much less than than the enclosed mass of the galaxy at r$\sim$40$\arcsec$, $\sim$~2.2~$\times$~10$^{9}$~M$_{\odot}$. That we see a star-forming ring is not conclusive evidence for a bar;  rings can be formed in a variety of ways including during accretion events or collisions/interactions between galaxies \citep[e.g.,][]{Berentzen03, Mapelli12}.  By themselves, our kinematic data do not suggest that a bar is currently present in NGC~6503.

While our data show no evidence for a bar in the kinematics (beyond 5$\arcsec$), a faint, but statistically significant bar-like feature \textit{is} detectable in the photometry.  To fully model the light distribution in NGC~6503, both a bar and a bulge are required.  Can this photometric feature consistently explain the $\sigma$-drop, Type II profile, and gas morphology given the regular kinematics?

Given the faintness of the photometric bar and the very well-behaved kinematic data, we find it is most likely that this bar is weak and it is the combination of the bar and the circumnuclear disc in NGC~6503 that work together to produce the $\sigma$-drop.  Without more stellar kinematic data, however, we cannot conclusively rule out the possibility that other mechanisms, as described above, are not responsible for creating the $\sigma$-drop.  

On the contrary, it is plausible that the observed Type II surface brightness profile could be caused by the photometric bar.  Consistent with the simulations of \citetalias{Bureau05}, the bar we detect could be strong enough to rearrange the material in the disc, producing the characteristic Type II profile.   The $i = 75\degr$ simulations of \citetalias{Bureau05} (see their figure 4) and an intrinsic axial ratio $q$~$\sim$~0.2 (Tables~\ref{datatable} and \ref{phottable}), appropriate for the disc geometry of NGC~6503, suggest that relatively weak bars may produce this feature because the low-density minor axis of the bar is exposed.  The bar also has to be very close to end-on (as the \df\ model suggests it is), otherwise the characteristic signatures in the surface brightness profile would be less pronounced and the galaxy isophotes would have more of a ``peanut"-shape rather than a ``bulge+ring"-shape (see the right-most column of figure 4 in \citetalias{Bureau05}).

It may be possible to explain the properties of NGC~6503 by invoking a strong bar which dissolved sometime in the past.  For this scenario to work in NGC~6503, the bar must have once been much stronger, so as to be able to rearrange the disc material, before significantly weakening, or nearly dissolving, so that we now see very little trace of it in either the kinematics or photometry.  According to \citetalias{Bureau05}, who also study the time evolution of the surface brightness and stellar velocity dispersion profiles, it is only late in the lifetime of the bar, well after it has formed and buckled, that the Type II plateau and central $\sigma$-drop become significant and well-established.  This would suggest that the bar in NGC~6503 is ``old".  It also suggests that the star-forming ring is not caused by the bar (see above for alternative ring formation mechanisms), as it would be puzzling for such a bar to now be responsible for the formation of a young (less than $\sim$~0.5~Gyr; \citetalias{Freeland10}) ring.  Estimating the decay timescale for a bar is a challenging numerical problem and different techniques and studies yield conflicting results, though the general conclusion is that bars \textit{can} weaken and/or disappear \citep[see, e.g.,][for a discussion]{Shen04, Kormendy04}.

In the absence of a full suite of models, we cannot conclusively determine the evolutionary history of NGC~6503 and unequivocally establish that a (strong) bar is responsible for the observed $\sigma$-drop, Type II surface brightness profile and star-forming ring.  We \textit{can} say, though,  that a strong bar is not currently present in the galaxy.  The bar that exists today in NGC~6503 is weak enough to avoid being detected in the gas kinematics, but is just strong enough to leave a measurable photometric signature.

%%%%%%%%%%%%%%%%%%%%%%%%%%%%
\section{Summary}
%%%%%%%%%%%%%%%%%%%%%%%%%%%%
\label{summary}

We have obtained new, multi-wavelength, high-quality spectroscopic and photometric data for the nearby spiral galaxy NGC~6503.  Specifically, we have presented two H$\alpha$ velocity fields obtained using the DensePak and SparsePak IFUs, a CO(J=1-0) velocity field obtained using CARMA, and a re-derived VLA \HI\ velocity field.  We have also presented a CFHT WIRCAM $K$s-band image of the galaxy.  

We have introduced the first public release of \df, a code that fits non-parametric models to either velocity fields or images.  The kinematic models implemented in \df\ fit for rotation, as well as $m=0,1,2$ non-circular flows.  \df\ is also able to fit for a symmetric outer disc warp.  In the photometric models, \df\ can fit up to 3 components: a  non-parametric disc and bar, as well as a S\'ersic bulge.  Relative to other algorithms,  \df\ excels at finding relatively weak, coherent asymmetries and providing a robust estimate of the underlying disc properties in their presence.

We have used \df\ to model our velocity field data and search for non-circular motions.  We find the velocity fields and derived rotation curves to all be well-described by rotation-only models.  We find no evidence for coherent radial, lopsided, or bisymmetric non-circular motions in any of the datasets, nor a symmetric warp in the \HI\ data.  We compare the best-fitting model parameters and rotation curves for each dataset and find them to be in good agreement with each other, suggesting that they are all tracing the same underlying disc kinematics.  We combine the H$\alpha$ and \HI\ rotation curves to determine a master rotation curve for NGC~6503.

We have also used \df\ to analyze the $K$s-band image of NGC~6503 and produce models of the disc, bar, and bulge.  We find the galaxy to be an axisymmetric, disc-dominated system that contains a bulge and weak end-on bar.  Together, the bulge and bar contribute only 5.3\% of the total light.  We find the galaxy to have a Type II surface brightness profile that could have been caused by the bar.

NGC~6503 is a disc galaxy with regular, well-behaved kinematics.  Despite indirect photometric indications of the presence of a bar (e.g., star-forming ring, Type II surface brightness profile) and the faint (0.7\% of the total light) detection in our $K$s-band photometry, we do not find kinematic counterparts to these features in the \df\ models of the velocity field data; at least at present, the bar is weak.  The large radial extent and gently declining outer rotation curve of the galaxy can place interesting constraints on the concentration of the dark matter halo \citep[e.g.][]{Casertano91}; NGC~6503 thus remains one of the best targets for detailed studies of the dark and baryonic structure of late-type disc galaxies. However, particularly in the event that the bar was once stronger and caused the Type II profile, to constrain the \textit{inner} dark matter halo, full modeling of a barred galaxy that follows the angular momentum exchange between the disc and halo is required.

%%%%%%%%%%%%%%%%%%%%%%%%%%%%
\section*{Acknowledgements}
R.K.D., C.A.A., \& K.S. acknowledge funding from the Natural Sciences and Engineering Research Council of Canada (NSERC).  M.M. was supported by NASA through SAO Award Number 2834-MIT- SAO-4018, which is issued by the Chandra X-ray Observatory Center on behalf of NASA under contract NAS8-03060.  We would like to thank the referee for helpful comments that have improved this paper.  
%%%%%%%%%%%%%%%%%%%%%%%%%%%%

%%%%%%%%%%%%%%%%%%%%%%%%%%%%

%%%%%%%%%%%%%%%%%%%%%%%%%%%%

%%%%%%%%%%%%%%%%%%%%%%%%%%%%
\appendix
\section{Master Rotation Curve}
In Table~\ref{rcdata} we present the data for the master rotation curve for NGC~6503 discussed in Section~\ref{Comparerc} and plotted in Figure~\ref{masterrc}.

\begin{table}
\begin{center}
\caption{Master Rotation Curve for NGC~6503.  \label{rcdata}}
\begin{tabular}{|ccc}
\hline
 R 				&V  				&$\sigma_{V}$\\ 
 ($\arcsec$)		&(km s$^{-1}$)		&(km s$^{-1}$)\\
\hline
       6.0      		&17.8        		&8.0\\
     12.0    		&30.9        		&5.4\\
     18.0     		&52.6        		&4.6\\
     24.0     		&72.8        		&3.8\\
     30.0     		&87.4        		&3.5\\
     36.0     		&95.6        		&3.6\\
     42.0     		&103.0        	&3.2\\
     48.0     		&105.4        	&3.0\\
     54.0     		&109.5        	&2.7\\
     60.0     		&109.7        	&2.6\\
     66.0     		&112.7        	&2.8\\
     72.0     		&112.8        	&2.6\\
     78.0     		&119.2        	&2.6\\
     84.0     		&123.5        	&2.8\\
     90.0     		&121.6        	&3.1\\
     96.0     		&124.3        	&2.8\\
    102.0     	&121.9        	&3.0\\
    108.0     	&124.6        	&2.9\\
    114.0     	&125.0        	&3.6\\
    120.0     	&125.9        	&3.8\\
    126.0     	&123.2        	&3.5\\
    132.0     	&119.9        	&3.8\\
    138.0     	&127.2        	&4.4\\
    144.0     	&123.4        	&6.3\\
    150.0     	&120.0 		&5.1\\
    165.0       	&123.6        	&1.9\\
    180.0      	&122.4        	&1.9\\
    195.0       	&120.8        	&1.7\\
    210.0       	&120.5        	&2.0\\
    225.0     	&120.8        	&2.5\\
    240.0       	&112.6        	&2.7\\
    255.0       	&118.6        	&2.7\\
    270.0       	&113.4        	&2.6\\
    285.0       	&116.2         	&2.5\\
    300.0       	&115.6        	&2.2\\
    315.0       	&115.3        	&2.4\\
    330.0       	&116.2        	&2.3\\
    345.0       	&116.8        	&2.5\\
    360.0       	&119.1        	&2.2\\
    375.0       	&115.4        	&2.4\\
    390.0         	&117.0        	&2.2\\
    405.0       	&114.3        	&2.2\\
    420.0       	&114.8         	&2.1\\
    435.0       	&117.1        	&2.1\\
    450.0       	&114.6        	&2.4\\
    465.0       	&116.3        	&2.2\\
    480.0       	&112.5        	&2.4\\
    495.0       	&114.3        	&2.1\\
    510.0         	&114.0        	&2.2\\
    525.0       	&114.9        	&2.0\\
    540.0         	&115.0        	&2.2\\
    555.0       	&115.8        	&2.4\\
    570.0       	&114.9         	&2.3\\
    585.0       	&114.2        	&2.8\\
    600.0         	&114.0        	&2.4\\
    630.0       	&113.1        	&2.6\\
    675.0       	&108.1        	&2.6\\
    720.0       	&105.6         	&3.1\\
    765.0       	&108.2        	&2.8\\
    810.0       	&102.8        	&2.1\\
\hline
\end{tabular}
\end{center}
\end{table}

\end{document}